\documentclass{LMCS}

\usepackage{enumerate}
\usepackage{hyperref}
\usepackage{graphicx}

\theoremstyle{plain}

\newcommand{\diam}{\mathop{\mathrm{diam}}}

\newcommand{\mesh}{\mathop{\mathrm{mesh}}}
\newcommand{\fdiam}{\mathop{\mathrm{fdiam}}}
\newcommand{\fmesh}{\mathop{\mathrm{fmesh}}}

\def\doi{7 (3:05) 2011}
\lmcsheading%
{\doi}
{1--21}
{}
{}
{Nov.~30, 2010}
{Aug.~25, 2011}
{}

\begin{document}

\title[Co-c.e.\ spheres and  cells  in computable metric spaces]{Co-c.e.\  spheres and  cells  in computable metric spaces}

\author[Z.~Iljazovi\'{c}]{Zvonko Iljazovi\'{c}}   
\address{Department of Mathematics, Faculty of Science, University of Zagreb, Bijeni\v{c}ka 30, 10 000 Zagreb, Croatia} 
\email{zilj@math.hr}  





\begin{abstract}
  \noindent    We investigate conditions under which a co-computably enumerable set in
  a computable metric space is computable. Using higher-dimensional chains and spherical chains
  we prove that in each computable metric space which is locally computable each
  co-computably enumerable sphere
  is computable and each co-c.e.\ cell with co-c.e.\ boundary sphere is
  computable.
\end{abstract}

\maketitle

\section{Introduction}\label{introd}
A closed subset of $\mathbb{R}^{m} $ is said to be
\textit{computable} if it can be effectively approximated by a
finite set of points with rational coordinates with arbitrary
given precision on arbitrary given bounded region of
$\mathbb{R}^{m} $. A closed subset of $\mathbb{R}^{m} $ is said to
be \textit{co-computable enumerable (co-c.e.)} if its complement
can be effectively covered by open balls. Each computable set is
co-c.e. On the other hand, there exist co-c.e.\ sets  which are
not computable. In fact, while each nonempty computable set
contains computable points, there exists a nonempty co-c.e.\ set
which contains no computable points (\cite{sp:mib}). Although the
implication
\begin{equation}\label{intro-2}
S\mbox{ co-computably enumerable } \Rightarrow S \mbox{
computable}
\end{equation}
does not hold in general, there are certain conditions under which
it does hold. The following result has been proved in
\cite{mi:mib}:
\begin{enumerate}[(1)]
\item[(i)] if $S\subseteq \mathbb{R}^{m} $ is homeomorphic to
$S^{n} $, where $S^{n} \subseteq \mathbb{R}^{n+1}$ is the unit
sphere,  then (\ref{intro-2}) holds; \item[(ii)] if $S\subseteq
\mathbb{R}^{m} $ is such that there exists a homeomorphism
$f:B^{n} \rightarrow S$, where $B^{n} \subseteq \mathbb{R}^{n} $
is the unit ball, such that $f(S^{n-1})$ is a co-c.e.\ set, then
(\ref{intro-2}) holds.
\end{enumerate}
In the case $n=1$, i.e$.$ in the case when $S$ is a topological
circle or when $S$ is a co-c.e$.$
 arc with computable endpoints, the preceding result
  has been generalized in \cite{zi:mib} to
 computable metric spaces with the effective covering property and
 compact closed balls.
 Furthermore, by \cite{zi2:mib},
 the assumption of the effective covering property and
 compact closed balls can be replaced here by the weaker assumption that
 a computable metric space is locally computable.

 In this paper we prove that this result holds for every $m\geq 1$, i.e. we prove that if $(X,d,\alpha )$ is a computable metric
 space which is locally computable, then
\begin{enumerate}[(1)]
\item[(i)] if $S$ is a co-c.e.\ set in $(X,d,\alpha )$
homeomorphic to $S^{n} $, then $S$ is computable in $(X,d,\alpha
)$; \item[(ii)] if $f:B^{n} \rightarrow S$ is a homeomorphism,
where $S$ is co-c.e.\ in $(X,d,\alpha )$, such that $f(S^{n-1})$
is also co-c.e., then $S$ is computable.
\end{enumerate}

In order to prove this, we use techniques similar to those in
\cite{zi:mib}. In Section \ref{sect-topology} we examine the
topological side of the problem. We define the notions of
$n-$chain and spherical $n-$chain in a metric space. These notions
play the same role as the notions of a chain and a circular chain
play in the proof of the main results of \cite{zi:mib}. However,
the higher-dimensional aspect of the problem will require  some
deeper topological facts and we will rely here on a result proved
in \cite{re:mib}. In Section \ref{sect-comp} we include
computability into consideration and we prove the main results of
the paper: Theorem \ref{thm-1} and Theorem \ref{thm-2}.

\section{Preliminaries}\label{prelim}
If $X$ is a set, let $\mathcal{P}(X)$ denote the set of all
subsets of $X$.

For $m\in \mathbb{N}$ let  $\mathbb{N}_{m} =\{0,\dots ,m\}$. For
$n\geq 1$ let $$\mathbb{N}_{m} ^{n} =\{(x_{1} ,\dots ,x_{n} )\mid
x_{1} ,\dots ,x_{n}\in \mathbb{N}_{m} \}.$$

We say that a function $\Phi :\mathbb{N}^{k} \rightarrow
\mathcal{P}(\mathbb{N}^{n} )$ is \textbf{computable} if the
function $\overline{\Phi }:\mathbb{N}^{k+n}\rightarrow \mathbb{N}$
defined by
$$\overline{\Phi }(x,y)=\chi _{\Phi (x)}(y),$$ $x\in \mathbb{N}^{k} ,$ $y\in \mathbb{N}^{n}$
is computable (i.e.\ recursive). Here  $\chi _{S}:\mathbb{N}^{n}
\rightarrow \{0,1\}$ denotes the characteristic function of
$S\subseteq \mathbb{N}^{n} $. A function $\Phi :\mathbb{N}^{k}
\rightarrow \mathcal{P}(\mathbb{N}^{n} )$ is said to be
\textbf{computably bounded} if there exists a computable function
$\varphi :\mathbb{N}^{k} \rightarrow \mathbb{N}$ such that $\Phi
(x)\subseteq \mathbb{N}_{\varphi (x)}^{n} $,  for all $x\in
\mathbb{N}^{k}$.

We say that a function $\Phi :\mathbb{N}^{k} \rightarrow
\mathcal{P}(\mathbb{N}^{n} )$ is  \textbf{c.c.b}$.$  if $\Phi $ is
computable and computably bounded.

\begin{prop} \label{p1}\hfill
\begin{enumerate}[\em(1)]
\item If $\Phi ,\Psi:\mathbb{N}^{k} \rightarrow
\mathcal{P}(\mathbb{N}^{n} )$ are c.c.b.\ functions, then the sets
$\{x\in \mathbb{N}^{k} \mid \Phi (x)=\Psi (x)\}$ and $\{x\in
\mathbb{N}^{k} \mid \Phi (x)\subseteq \Psi (x)\}$ are decidable.

\item Let $\Phi :\mathbb{N}^{k} \rightarrow
\mathcal{P}(\mathbb{N}^{n} )$ and $\Psi :\mathbb{N}^{n}\rightarrow
\mathcal{P}(\mathbb{N}^{m} )$ be c.c.b.\ functions. Let $\Lambda
:\mathbb{N}^{k} \rightarrow \mathcal{P}(\mathbb{N}^{m} )$ be
defined by $$\Lambda (x)=\bigcup_{z\in \Phi (x)}\Psi  (z),$$ $x\in
\mathbb{N}^{k} $. Then $\Lambda  $ is a c.c.b.\ function.

\item Let $\Phi :\mathbb{N}^{k} \rightarrow
\mathcal{P}(\mathbb{N}^{n} )$ be c.c.b$.$ and let $T\subseteq
\mathbb{N}^{n} $ be c.e. Then the set $S=\{x\in \mathbb{N}^{k}
\mid \Phi (x)\subseteq T\}$ is c.e. \qed
\end{enumerate}
\end{prop}

\noindent A function $F:\mathbb{N}^{k+1} \rightarrow \mathbb{Q}$ is called
\textbf{computable} if there exist computable functions
$a,b,c:\mathbb{N}^{k+1}\rightarrow \mathbb{N}$ such that
$$F(x)=(-1)^{c(x)}\frac{a(x)}{b(x)+1}$$ for each $x\in
\mathbb{N}^{k+1} $. A number $x\in \mathbb{R}$ is said to be
\textbf{computable} if there exists a computable function
$g:\mathbb{N}\rightarrow \mathbb{Q}$ such that $|x-g(i)|<2^{-i}$
for each $i\in \mathbb{N}$.

By a \textbf{computable} function $\mathbb{N} ^{k} \rightarrow
\mathbb{R}$ we mean a function $f:\mathbb{N} ^{k} \rightarrow
\mathbb{R}$ for which there exists a computable function
$F:\mathbb{N}^{k+1}\rightarrow \mathbb{Q}$ such that
$$|f(x)-F(x,i)|<2^{-i}$$ for all $x\in \mathbb{N}^{k}$ and $i\in
\mathbb{N}$.

In the following proposition we state some elementary facts about
computable functions $\mathbb{N}^{k} \rightarrow \mathbb{R}.$
\begin{prop} \label{NuR}\hfill
\begin{enumerate}[\em(1)]

\item If $f,g:\mathbb{N}^{k} \rightarrow \mathbb{R}$ are
computable, then $f+g,f-g:\mathbb{N}^{k} \rightarrow \mathbb{R}$
are computable.

\item If $f:\mathbb{N}^{k} \rightarrow \mathbb{R}$ and
$F:\mathbb{N}^{k+1} \rightarrow \mathbb{R}$ are functions such
that $F$ is computable and $|f(x)-F(x,i)|<2^{-i},$ $\forall x\in
\mathbb{N}^{k} ,$ $\forall i\in \mathbb{N},$ then $f$ is
computable.

\item If $f:\mathbb{N}^{n+1}\rightarrow \mathbb{R}$ and $\varphi
:\mathbb{N}\rightarrow \mathbb{N}$ are computable functions, then
the function $g:\mathbb{N} \rightarrow \mathbb{R}$ defined by
$$g(l)=\max_{0\leq j_{1} ,\dots ,j_{n} \leq \varphi (l)}f(l,j_{1} ,\dots ,j_{n} )$$
is computable.

\item If $f,g:\mathbb{N}^{k} \rightarrow \mathbb{R}$ are
computable functions, then the set $\{x\in \mathbb{N}^{k} \mid
f(x)>g(x)\}$ is c.e.
\end{enumerate} \qed
\end{prop}

\noindent A tuple $(X,d,\alpha )$ is said to be a \textbf{computable metric
space} if $(X,d)$ is a metric space and $\alpha :\mathbb{N}
\rightarrow X$ is a sequence dense in $(X,d)$ (i.e. a sequence
which range is dense in $(X,d)$) such that the function
$\mathbb{N} ^{2}\rightarrow \mathbb{R} $, $$(i,j)\mapsto
d(\alpha_{i} ,\alpha_{j} )$$ is computable (we use notation
$\alpha =(\alpha _{i} )$).

If  $(X,d,\alpha )$ is a computable metric space, then a sequence
$(x_{i} )$ in $X$ is said to be \textbf{computable} in
$(X,d,\alpha )$ if there exists a computable function
$F:\mathbb{N} ^{2}\rightarrow \mathbb{N} $ such that $$d(x_{i}
,\alpha _{F(i,k)})<2^{-k}$$ for all $i,k\in \mathbb{N} $. A point
$a \in X$ is said to be \textbf{computable} in $(X,d,\alpha )$ if
the constant sequence $a ,a ,\dots $ is computable.

Let $(X,d,\alpha )$ be a computable metric space. Let
$q:\mathbb{N}\rightarrow \mathbb{Q}$ be some fixed computable
function whose image is $\mathbb{Q}\cap \langle 0,\infty\rangle $
and let $\tau ,\tau' :\mathbb{N}\rightarrow \mathbb{N}$ be some
fixed computable functions such that $\{(\tau (i),\tau' (i))\mid
i\in \mathbb{N}\}=\mathbb{N}^{2}.$ For $i\in \mathbb{N}$ we define
$$I_{i}=B(\alpha_{\tau (i) } ,q_{\tau '(i)} ),~\widehat{I}_{i}=\widehat{B}(\alpha_{\tau (i)}
,q_{\tau '(i)} ).$$ Here, for $x\in X$ and $r>0$, we denote by
$B(x,r)$ the open ball of radius $r$ centered at $x$ and by
$\widehat{B}(x,r)$ the corresponding closed ball, i.e$.$
$B(x,r)=\{y\in X\mid d(x,y)<r\},$ $\widehat{B}(x,r)=\{y\in X\mid
d(x,y)\leq r\}$.  For $A\subseteq X$ we will denote the closure of
$A$ by $\overline{A}$.

As a consequence of Proposition \ref{NuR} we get the following
corollary.
\begin{cor} \label{kIi}
Let $(X,d,\alpha )$ be a computable metric space. The set
$\{(k,i)\in \mathbb{N}^{2}\mid \alpha _{k} \in I_{i}\}$ is c.e.
\qed
\end{cor}

A closed  subset $S$ of $(X,d)$ is said to be \textbf{computably
enumerable} in $(X,d,\alpha )$ if $$\{i\in \mathbb{N}\mid S \cap
I_{i} \neq\emptyset \}$$ is a c.e$.$ subset of $\mathbb{N}.$ A
closed subset $S$ is said to be \textbf{co-computably enumerable}
in $(X,d,\alpha )$ if there exists a computable function
$f:\mathbb{N}\rightarrow \mathbb{N}$ such that
$$X\setminus S=\bigcup _{i\in \mathbb{N}}I_{f(i)}.$$ It is easy to
see that these definitions do not depend on functions $\tau ,$
$\tau' $ and $q$. We say that $S$ is a \textbf{computable} set in
$(X,d,\alpha )$ if $S$ is both computably enumerable and
co-computably enumerable (\cite{bp:mib,we}).

Let $\sigma :\mathbb{N}^{2}\rightarrow \mathbb{N}$ and
$\eta:\mathbb{N}\rightarrow \mathbb{N}$ be some fixed computable
functions with the following property: $\{(\sigma (j,0),\dots
,\sigma (j,\eta(j)))\mid j\in \mathbb{N}\}$ is the set of all
finite sequences in $\mathbb{N}$ excluding the empty sequence,
i.e. the set $\{(a_{0} ,\dots ,a_{n} )\mid n\in  \mathbb{N},~a_{0}
,\dots ,a_{n} \in \mathbb{N}\}.$ Such functions, for instance, can
be defined using the Cantor pairing function. We use the following
notation: $(j)_{i}$ instead of $\sigma (j,i)$ and $\overline{j}$
instead of $\eta(j).$  Hence $$\{((j)_{0} ,\dots
,(j)_{\overline{j}})\mid j\in \mathbb{N}\}$$ is the set of all
finite sequences in $\mathbb{N}.$ For $j\in \mathbb{N}$ let $[j]$
be defined by
\begin{equation}\label{p2-eq}
[j]=\{(j)_{i} \mid 0\leq i\leq \overline{j}\}.
\end{equation}
Note that the function $\mathbb{N}\rightarrow
\mathcal{P}(\mathbb{N})$, $j\mapsto [j]$, is c.c.b.

Let $(X,d,\alpha )$ be a computable metric space.  For $j\in
\mathbb{N}$ we define
$$J_{j}=\bigcup_{i\in [j]}I_{i},~~\widehat{J}_{j}=\bigcup _{i\in [j]}\widehat{I}_{i}.$$
The sets $J_{j} $ represent finite unions of  rational balls and
the sets $\widehat{J}_{j}$ finite unions of  closed rational
balls.

\begin{cor} \label{kJj}
Let $(X,d,\alpha )$ be a computable metric space. The set
$\{(k,j)\in \mathbb{N}^{2}\mid \alpha _{k} \in J_{j}\}$ is c.e.
\end{cor}
\proof We have $\alpha _{k} \in J_{j}$ if and only if there exists
$i\in \mathbb{N}$ such that $i\leq \overline{j}$ and $\alpha
_{k}\in I_{(j)_{i} }$ and the claim follows from Corollary
\ref{kIi}. \qed

A computable metric space $(X,d,\alpha )$ has the
\textbf{effective covering property} if the set $$\{(w,j)\in
\mathbb{N}^{2}\mid \widehat{I}_{w}\subseteq J_{j}\}$$ is
computably enumerable (\cite{bp:mib}). It is not hard to see that
this definition does not depend on the choice of the functions
$q,\tau ,\tau ' ,\sigma ,\eta$ which are necessary in the
definitions of sets ${I}_{w}$ and $J_{j}.$

For example, if $\alpha :\mathbb{N} \rightarrow \mathbb{R}^{n}  $
is a computable function (in the sense that the component
functions of $\alpha $ are computable) whose image is dense in
$\mathbb{R}^{n} $ and $d$ is the Euclidean metric on $\mathbb{R}$,
then $(\mathbb{R}^{n}  , d, \alpha )$ is a computable metric
space. A sequence $(x_{i} )$ is computable in this computable
metric space if and only if $(x_{i} )$ is a computable sequence in
$\mathbb{R}^{n} $  and $(x_{1} ,\dots ,x_{n} )\in \mathbb{R}^{n} $
is a computable point in this space if and only if $x_{1} $,\dots
,$x_{n} $ are computable numbers. This computable metric space has
the effective covering property (see e.g.\ \cite{zi:mib}).

If $(X,d,\alpha )$ is a computable metric space, then a compact
set $K$ in $(X,d)$ is said to be \textbf{computably compact} in
$(X,d,\alpha )$ if $K$ is computably enumerable in $(X,d,\alpha )$
and if the set $\{j\in \mathbb{N}\mid K\subseteq J_{j} \}$  is
c.e.\ (\cite{br:mib}). A computable metric space $(X,d,\alpha )$
is \textbf{locally computable}  (\cite{br:mib}) if for each
compact set $A$ in $(X,d)$ there exists a computably compact set
$K$ in $(X,d,\alpha )$ such that $A\subseteq K$.

Let $(X,d,\alpha )$ be a computable metric space. A computable
metric space $(Y,d',\beta )$ is said to be a \textbf{subspace} of
$(X,d,\alpha )$ if $Y\subseteq X$, $d':Y\times Y\rightarrow
\mathbb{R}$ is the restriction of $d:X\times X\rightarrow
\mathbb{R}$ and $\beta $ is a computable sequence in $(X,d,\alpha
)$.

The proofs of  the following propositions can be found in
\cite{zi2:mib}.
\begin{prop} \label{potpr-zi2}
Let $(Y,d',\beta )$ be a subspace of a computable metric space
$(X,d,\alpha )$ and let $S\subseteq Y$.
\begin{enumerate}[\em(1)]

\item[(i)] If $S$ is co-c.e.\ in $(X,d,\alpha )$, then $S$ is
co-c.e.\ in $(Y,d',\beta )$.

\item[(ii)] If $S$ is c.e.\ in $(X,d,\alpha )$, then $S$ is c.e.\
in $(Y,d',\beta )$. Conversely, if $S$ is closed in $(X,d)$ and
c.e.\ in $(Y,d',\beta  )$, then $S$ is c.e.\ in $(X,d,\alpha )$.
\qed
\end{enumerate}
\end{prop}
\begin{prop} \label{rekomp-pot}
Let $(X,d,\alpha )$ be a computable metric space and let $K$ be a
nonempty compact set in $(X,d)$. Then $K$ is computably compact in
$(X,d,\alpha )$ if and only if there exist a metric $d'$ on $K$
and a sequence $\beta $ in $K$ such that $(K,d',\beta )$ is a
subspace of $(X,d,\alpha )$ and $(K,d',\beta )$ has the effective
covering property. \qed
\end{prop}

\section{$n-$chains and spherical $n-$chains}\label{sect-topology}

For $n\geq 1$ let
$$B^{n}=\{x\in \mathbb{R}^{n} \mid \|x\|\leq
 1\}$$ and
 $$S^{n-1}=\{x\in \mathbb{R}^{n} \mid \|x\|=
 1\}.$$
 A topological space $X$ is called an $n-$\textbf{cell} if it is
 homeomorphic to $B^{n}$. We say that $X$ is an $n-$\textbf{sphere} if it is
 homeomorphic to $S^{n} $.

  By the \textbf{boundary sphere} of an $n-$cell $E$
  we mean the set $f(S^{n-1})$, where $f:B^{n} \rightarrow E$
 is a homeomorphism. (Note that the boundary sphere of $E$, when $E$ is a subspace of some topological space $X$,
 need not be equal to the topological
 boundary of $E$ in $X$.) The definition of the boundary sphere does not depend
 on a particular homeomorphism $f:B^{n} \rightarrow E$.
 Namely, this a consequence of the fact that each homeomorphism
 $B^{n} \rightarrow B^{n} $
 maps $S^{n-1}$ onto $S^{n-1}$ (or equivalently $B^{n} \setminus S^{n-1}$
  onto $B^{n} \setminus S^{n-1}$)  which follows from the
  Invariance of domain theorem (see \cite{mu}): if $h:U\rightarrow
  \mathbb{R}^{n} $ is continuous and injective, where $U$ is an
  open subset of $\mathbb{R}^{n} $, then $h(U)$ is open.

The result that we want to prove can now be restated in this way:
if $(X,d,\alpha )$ is a computable metric space which is locally
computable, then
\begin{enumerate}[(1)]
\item[(1)] each co-c.e.\ $n-$sphere is computable; \item[(2)] each
co-c.e.\ $n-$cell whose boundary sphere is co-c.e.\ is computable.
\end{enumerate}

Let us first note that it is enough to prove this result in the
case when $(X,d,\alpha )$ is a computable metric space which has
the effective covering property and compact closed balls. Namely,
suppose that the result holds for such computable metric spaces
and let $(X,d,\alpha )$ be a computable metric space which is
locally computable. Let $S\subseteq X$ be a co-c.e.\ $n$-sphere.
Then $S\subseteq K$, where $K$ is computably compact in
$(X,d,\alpha )$. By Proposition \ref{rekomp-pot} there exist $d'$
and $\beta $ such that $(K,d',\beta )$ is a subspace of
$(X,d,\alpha )$ and such that $(K,d',\beta )$ has the effective
covering property. By Proposition \ref{potpr-zi2}(i) $S$ is
co-c.e.\ in $(K,d',\beta )$ and therefore $S$ is computable in
$(K,d',\beta )$. Proposition \ref{potpr-zi2}(ii) implies now that
$S$ is c.e.\ in $(X,d,\alpha )$, hence $S$ is computable in
$(X,d,\alpha )$. In the same way we get that each co-c.e.\
$n-$cell in $(X,d,\alpha )$ whose boundary sphere is co-c.e.\ is
computable.

Let us observe how the statement (2) was proved in \cite{zi:mib}
in the case $n=1$.  Let $E$ be a co-c.e.\ arc with computable
endpoints $a$ and $b$.  For each $\varepsilon >0$ there exists a
finite sequence of open sets $C_{0} ,\dots ,C_{m}$ such that
\begin{enumerate}[(1)]
\item[(i)] $E\subseteq C_{0} \cup \dots \cup C_{m} ;$

\item[(ii)] $a\in C_{0} $, $b\in C_{m} $;

\item[(iii)] $C_{i} \cap C_{j} =\emptyset$ for all $i,j$ such that
$|i-j|> 1$;

\item[(iv)] each $C_{i} $ is the finite union of rational balls,
i.e.\ it is equal to some $J_{j} $;

\item[(v)] $\diam C_{i} <\varepsilon$,
\end{enumerate}
where $\diam C_{i}$ denotes the diameter of the set $C_{i} $ (see
Figure 1). Since $S$ is co-c.e.\ and $(X,d,\alpha )$ has the
effective covering property and compact closed balls, it is
possible to find effectively for each $k\in \mathbb{N}$ sets
$C_{0} ,\dots ,C_{m} $ with properties (i)--(v), where
$\varepsilon =2^{-k}$.

\begin{center}
  \includegraphics[height=1.8in,width=2.7in]{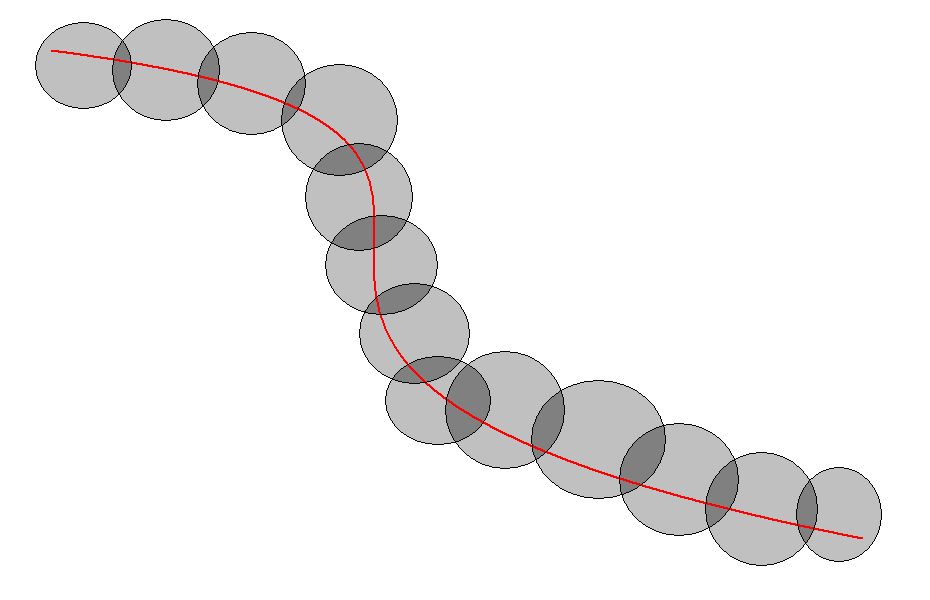}
  \includegraphics[height=1.8in,width=2.7in]{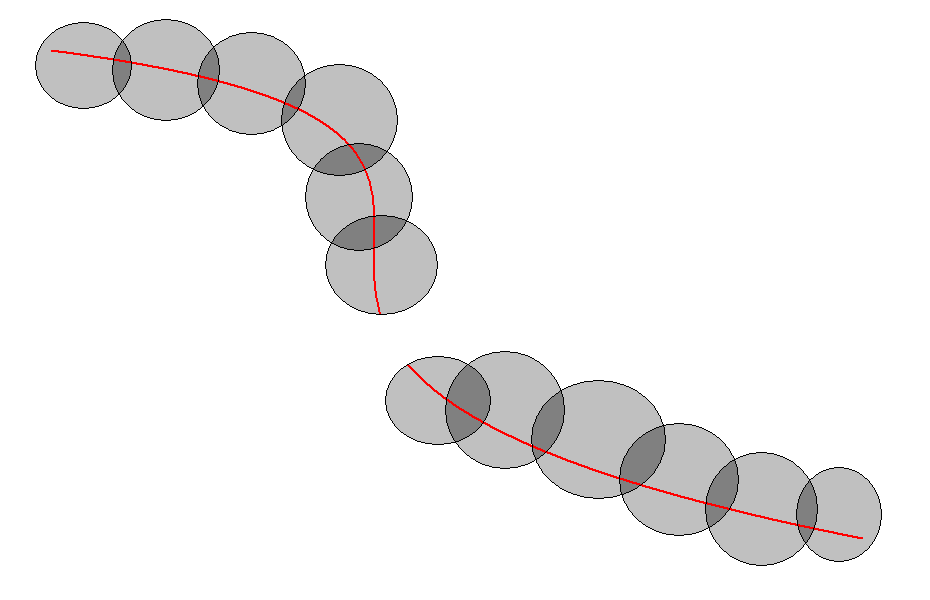}

  \emph{Figure 1.} \hspace{145pt} \emph{Figure 2.}
\end{center}

\noindent However, this means that we can effectively approximate $S$,
namely properties (i)--(v) imply that $C_{0} \cup \dots \cup C_{m}
$ is a $2^{-k}-$approximation of $S$ in the following sense: for
each $x\in E$ there exists $y\in C_{0} \cup \dots \cup C_{m}$ such
that $d(x,y)<2^{-k}$ and for each $y\in C_{0} \cup \dots \cup
C_{m}$ there exists $x\in E$ such that $d(y,x)<2^{-k}$. Using this
fact we can  prove that $E$ is computable.

Why properties (i)--(v) imply that $C_{0} \cup \dots \cup C_{m} $
is an $2^{-k}-$approximation of $S$? The fact that for each $x\in
E$ there exists $y\in C_{0} \cup \dots \cup C_{m}$ such that
$d(x,y)<2^{-k}$ follows trivially from (i). On the other hand, the
fact that for each $y\in C_{0} \cup \dots \cup C_{m}$ there exists
$x\in E$ such that $d(y,x)<2^{-k}$ can be easily deduced from (v)
and the fact that
\begin{equation}\label{CsijeceS}
C_{i} \cap S\neq \emptyset \mbox{ for each } i\in \{0,\dots ,m\}.
\end{equation}
But why (\ref{CsijeceS}) holds? If we assume  $C_{i} \cap
E=\emptyset $ for some $i\in \{0,\dots ,m\}$, then $0<i<m$ and
$C_{0} \cup \dots \cup C_{i-1}$ and $C_{i+1} \cup \dots \cup
C_{m}$ are two disjoint open sets (Figure 2.) which cover $E$ and
each of them intersects $E$ which contradicts the fact that $E$ is
connected.

Suppose now that $E$ is a $2-$cell which is co-c.e.\ and whose
boundary sphere is co-c.e. In order to prove that $E$ is
computable, we would like to proceed similarly as in the case of
an arc. Naturally,  in this case we are trying to find sets
$C_{i,j}$, $0\leq i,j\leq m$, which satisfy properties similar to
properties (i)--(v) with basic difference that instead of (iii) we
require
\begin{equation}\label{intro-6}
C_{i,j}\cap C_{i',j'}=\emptyset\mbox{ if } |i-i'|> 1\mbox{ or
}|j-j'|> 1.
\end{equation}
\begin{center}
  \includegraphics[height=1.8in,width=1.8in]{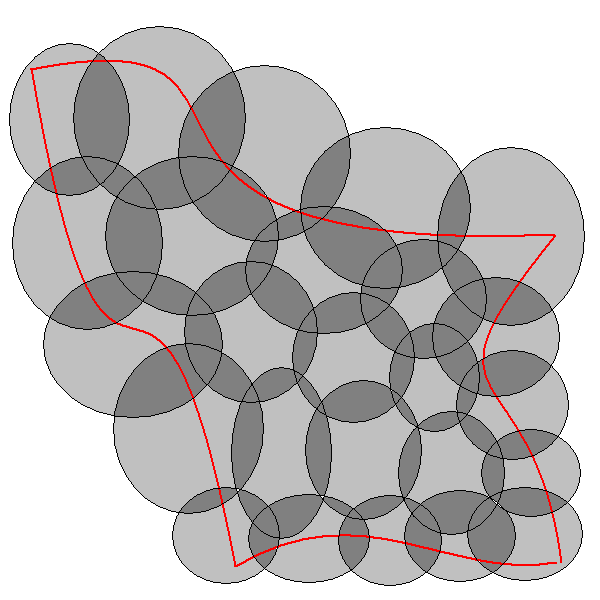}

  \emph{Figure 3.} The sets $C_{i,j}$ cover the 2-cell whose
  boundary sphere is the red curve
\end{center}
The main question here is what other properties we should require
so that those properties imply
\begin{equation}\label{CsijeceS2}
C_{i,j} \cap E\neq \emptyset \mbox{ for all } i,j\in \{0,\dots
,m\};
\end{equation}
the fact (\ref{CsijeceS2}) is important since we want to conclude
that $\bigcup_{i,j}C_{i,j}$ approximates $E$ in the same way as in
the case of an arc.

If we suppose that $i_{0} $ and $j_{0} $ are such that $0<i_{0}
<m$, $0<j_{0} <m$ and such that $C_{i_{0} ,j_{0} }\cap
E=\emptyset$, then we cannot conclude in general that $E$ is
covered by two disjoint open sets as in the case of an arc, but we
can define the sets
$$U=\bigcup_{i<i_{0} }\bigcup_{j}C_{i,j},~U'=\bigcup_{i>i_{0} }\bigcup_{j}C_{i,j},$$
$$V=\bigcup_{j<j_{0}
}\bigcup_{i}C_{i,j},~V'=\bigcup_{j>j_{0} }\bigcup_{i}C_{i,j},$$
and then these sets cover $E$ and we have $U\cap U'=\emptyset $,
$V\cap V'=\emptyset $. (See Figure 4.  The missing set is
$C_{i_{0} ,j_{0} }$. The
  vertical blue sets are $U$ and $U'$, the horizontal blue sets
  are $V$ and $V'$.)

\begin{center}
  \includegraphics[height=1.5in,width=1.5in]{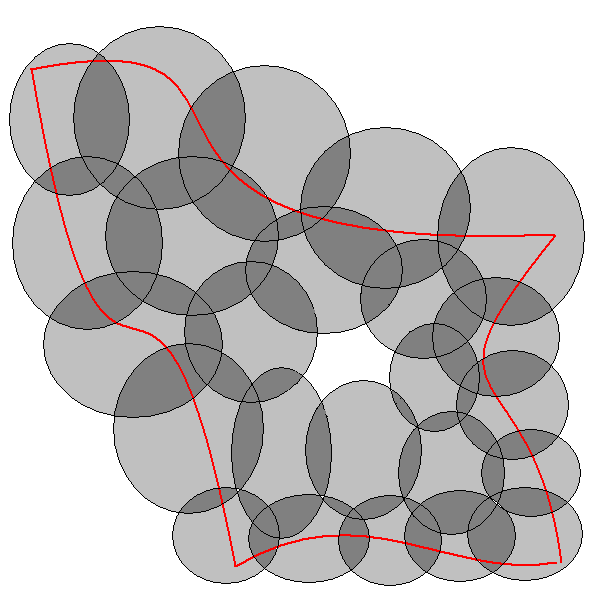}\hspace{20pt}
  \includegraphics[height=1.5in,width=1.5in]{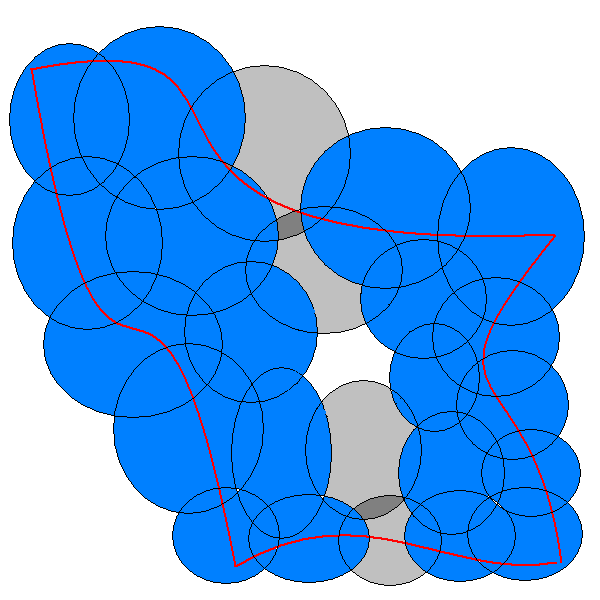}\hspace{20pt}
  \includegraphics[height=1.5in,width=1.5in]{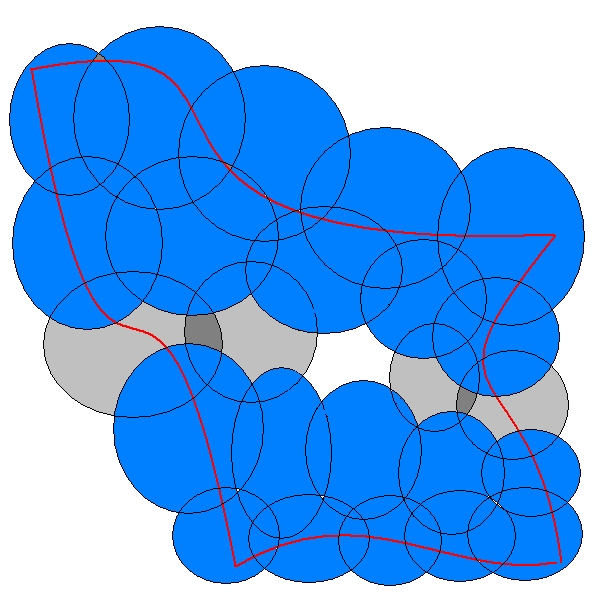}

  \emph{Figure 4.}
\end{center}

\noindent Since $E$ is homeomorphic to $I^{2}=[0,1]\times [0,1]$, this
raises the following question: is it possible to cover $I^{2}$ by
open sets $U$, $U'$, $V$ and $V'$ so that $U\cap U'=\emptyset $,
$V\cap V'=\emptyset $ and so that (see Figure 5.) $$\{0\}\times
[0,1]\subseteq U,~ \{1\}\times [0,1]\subseteq U',~ [0,1]\times
\{0\}\subseteq V,~ [0,1]\times \{1\}\subseteq V'?$$
\begin{center}
  \includegraphics[height=1.3in,width=1.5in]{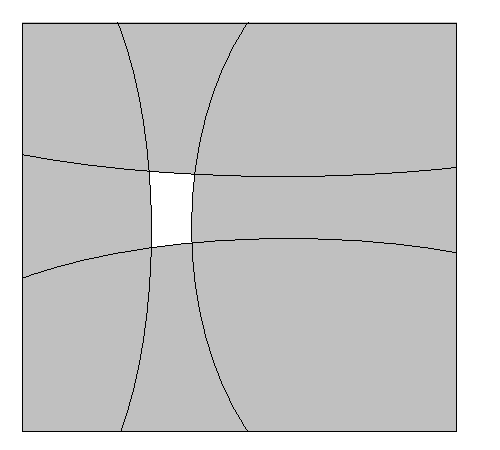}

  \emph{Figure 5.}
\end{center}

\noindent Let $X$ be a topological space and let $A$, $B$, $L$ be
subsets of $X$. We say that $L$ is a \textbf{partition between} $A$
and $B$ (see \cite{re:mib}) if there exist open sets $U$ and $W$ in
$X$ such that
$$A\subseteq U,~B\subseteq W,~U\cap W=\emptyset \mbox{ and }X\setminus
L=U\cup W.$$ For $n\geq 1$ let $$I^{n} =\{(x_{1} ,\dots ,x_{n}
)\in \mathbb{R}^{n} \mid x_{1} ,\dots ,x_{n} \in [0,1]\}.$$ For
$n\geq 1$ and $i\in \{1,\dots ,n\}$ let $$A^{n,0} _{i} =\{(x_{1}
,\dots ,x_{n} )\in I^{n} \mid x_{i} =0\},$$ $$A^{n,1} _{i}
=\{(x_{1} ,\dots ,x_{n} )\in I^{n} \mid x_{i} =1\}.$$ When the
context is clear, we write $A^{0}_{i} $ and $A^{1}_{i} $ instead
of $A^{n,0}_{i} $ and $A^{n,1}_{i} $. Let $\partial I^{n} $ denote
the boundary of $I^{n} $ in $\mathbb{R}^{n} $, hence
$$\partial I^{n} =A^{0}_{1}\cup \dots \cup A^{0}_{n} \cup A^{1} _{1} \cup \dots
\cup A^{1} _{n} .$$ It is a well known fact that there is a
homeomorphism $h:B^{n} \rightarrow I^{n} $ such that
$h(S^{n-1})=\partial I^{n} $. Hence if $E$ is an $n-$cell, then
there is a homeomorphism $f:I^{n} \rightarrow E$. In this case
$f(\partial I^{n} )$ is the boundary sphere of $E$.

The following theorem can be found in \cite{re:mib} (Theorem
1.8.1).
\begin{thm}\label{re-dimth}
Let $n\geq 1$. If $L_{i}$ is a partition between $A^{0}_{i} $ and
$A^{1}_{i} $ in $I^{n} $ for $i\in \{1,\dots ,n\}$, then
$\bigcap_{i=1}^{n} L_{i} \neq\emptyset $. \qed
\end{thm}

\begin{cor}\label{glavni} Let $n\geq 1$. Suppose $U_{1} ,\dots ,U_{n} $
and $V_{1} ,\dots ,V_{n} $ are open subsets of $I^{n} $ such that
$$U_{i} \cap A_{i} ^{1}=\emptyset ,~V_{i} \cap A_{i} ^{0}=\emptyset
\mbox{ and }U_{i} \cap V_{i} =\emptyset $$ for all $i\in
\{1,\dots,n\}$. Then $I^{n} \neq U_{1} \cup \dots \cup U_{n} \cup
V_{1} \cup \dots \cup V_{n}  $.
\end{cor}
\proof Suppose the opposite. Then $\{U_{1} ,\dots ,U_{n} ,V_{1}
,\dots,V_{n} \}$ is an open cover of $I^{n} $ and let $\lambda $
be its Lebesgue number. We can certainly find finitely many closed
subsets $B_{1} ,\dots ,B_{l} $ of $I^{n} $ whose union is $I^{n} $
and each of which has the diameter less than $\lambda $. Then each
of the sets $B_{1} ,\dots ,B_{l} $ is contained in some of the
sets $U_{1} ,\dots ,U_{n} ,V_{1} ,\dots ,V_{n} $.

For $i\in \{1,\dots ,n\}$ we define $F^{0}_{i}$ to be the union of
$A^{0}_{i} $ and all sets $B_{1} ,\dots ,B_{l} $ which are subsets
of $U_{i} $ and $F_{i} ^{1}$ to be the union of $A^{1}_{i} $ all
$B_{1} ,\dots ,B_{l} $ which are subsets of $V_{i} $. Then $F_{1}
^{0},\dots ,F_{n} ^{0},F_{1} ^{1},\dots ,F_{n} ^{1}$ are closed
subsets of $I^{n} $, their union is $I^{n} $ and for each $i\in
\{1,\dots ,n\}$ we have
$$A_{i} ^{0}\subseteq  F_{i} ^{0},~A_{i}
^{1}\subseteq F_{i} ^{1}\mbox{ and }F_{i} ^{0}\cap F_{i}
^{1}=\emptyset . $$ Let $i\in \{1,\dots ,n\}$. Since $F_{i} ^{0}$
and $F_{i} ^{1}$ are closed and disjoint, there exist open sets
$W_{i} ^{0}$ and $W_{i} ^{1}$ in $I^{n} $ which are disjoint and
such that $F_{i} ^{0}\subseteq W_{i} ^{0}$, $F_{i} ^{1}\subseteq
W_{i} ^{1}$. Let $L_{i} =I^{n} \setminus (W_{i} ^{0}\cup W_{i}
^{1})$. Then $L_{i} $ is a partition between $A_{i} ^{0}$ and
$A_{i} ^{1}$. We have
$$\bigcap_{i=1}^{n} L_{i} =I^{n} \setminus \bigcup_{i=1}^{n}
(W_{i} ^{0}\cup W_{i} ^{1})\subseteq I^{n} \setminus
\bigcup_{i=1}^{n} (F_{i} ^{0}\cup F_{i} ^{1})=\emptyset,$$ which
is impossible by Theorem \ref{re-dimth}. \qed

\begin{cor}\label{kor-glavni}
Let $n\geq 2$. Suppose $U_{1} ,\dots ,U_{n-1} $ and $V_{1} ,\dots
,V_{n-1} $ are open subsets of $\partial I^{n} $ such that
$$U_{i} \cap \left(A_{i} ^{n,1}\cap A_{n} ^{n,0}\right)=\emptyset ,~V_{i} \cap \left(A_{i} ^{n,0}\cap A_{n} ^{n,0}\right)=\emptyset
\mbox{ and }U_{i} \cap V_{i} =\emptyset $$ for all $i\in
\{1,\dots,n-1\}$. Let $E$ be the union of all $A_{i} ^{\rho }$
such that $1\leq i\leq n$, $\rho \in \{0,1\}$, $(i,\rho )\neq
(n,0)$, i.e$.$ $$E=A_{1} ^{n,0}\cup \dots \cup A_{n-1}^{n,0}\cup
A_{1}^{n,1}\cup \dots \cup A_{n} ^{n,1}.$$ Then $E$ is not
contained in the union $U_{1} \cup \dots \cup U_{n-1} \cup V_{1}
\cup \dots \cup V_{n-1} $.
\begin{center}
  \includegraphics[height=1.3in,width=1.5in]{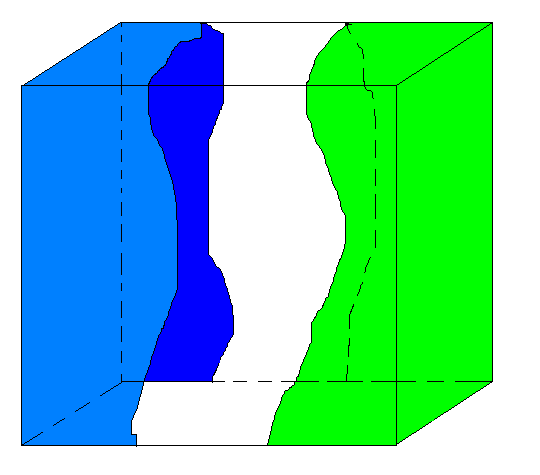}\hspace{25pt}
  \includegraphics[height=1.3in,width=1.5in]{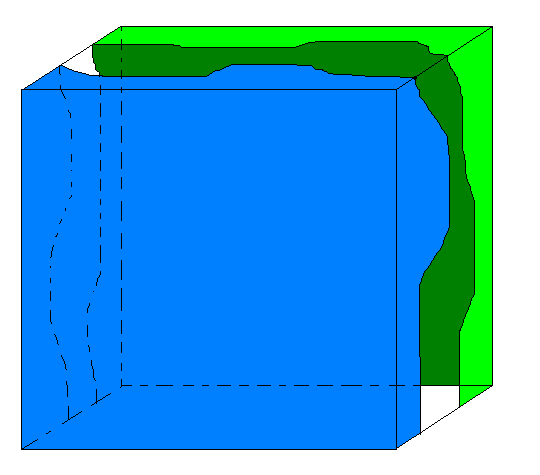}

  \emph{Figure 6.}
\end{center}

\end{cor}
\proof See Figure 6.\ (case $n=3$): the left and right blue sets
are $U_{1} $ and $U_{2} $
 respectively, and the left and right green sets are $V_{1} $ and $V_{2}
$ respectively. In this figure $E$ equals the union of vertical
faces of the cube and the upper face of the cube.

Let $f:E\rightarrow I^{n-1}$ be defined by
$$f(x_{1} ,\dots ,x_{n} )=\left(\frac{1}{2}+\frac{1}{2x_{n}
+1}\left(x_{1} -\frac{1}{2}\right),\dots
,\frac{1}{2}+\frac{1}{2x_{n} +1}\left(x_{n-1}
-\frac{1}{2}\right)\right).$$ It is straightforward to check  that
$f$ is bijective. Since $E$ is compact and $f$ clearly continuous,
$f$ is a homeomorphism. For each $i\in \{1,..,n-1\}$ we have
$$f\left(A_{i} ^{n,0}\cap A_{n} ^{n,0}\right)=A_{i} ^{n-1,0},~f\left(A_{i}
^{n,1}\cap A_{n} ^{n,0}\right)=A_{i} ^{n-1,1}.$$

Suppose that $E\subseteq U_{1} \cup \dots \cup U_{n-1} \cup V_{1}
\cup \dots \cup V_{n-1} $.  Then $$I^{n-1}=f(E\cap U_{1}
)\cup\dots \cup f(E\cap U_{n-1} )\cup f(E\cap V_{1} )\cup \dots
\cup f(E\cap V_{n} ).$$ For each $i\in \{1,\dots ,n-1\}$ the sets
$f(E\cap U_{i} )$ and $f(E\cap V_{i} )$ are open in $I^{n-1}$,
disjoint and
$$f(E\cap U_{i} )\cap A_{i} ^{n-1,1}=\emptyset ,~f(E\cap V_{i} )\cap A_{i}
^{n-1,0}=\emptyset.$$ This is impossible by Corollary
\ref{glavni}. \qed

For $i\in \{1,\dots ,n\}$ let
$$\partial^{0} _{i}\mathbb{N}_{m} ^{n} =\{(x_{1} ,\dots ,x_{n} )\in \mathbb{N}_{m} ^{n}
\mid x_{i} =0\},~\partial^{1} _{i}\mathbb{N}_{m} ^{n} =\{(x_{1}
,\dots ,x_{n} )\in \mathbb{N}_{m} ^{n} \mid x_{i} =m\}$$ and let
$$\partial \mathbb{N}_{m} ^{n} =\left(\bigcup_{1\leq i\leq
n}\partial_{i} ^{0}\mathbb{N}_{m} ^{n}\right)\cup
\left(\bigcup_{1\leq i\leq n}\partial_{i} ^{1}\mathbb{N}_{m}
^{n}\right).$$

Let $X$ be a set, $n\geq 1$ and $m\in \mathbb{N}$. A function
$$C:\mathbb{N}_{m} ^{n}  \rightarrow \mathcal{P}(X)$$  is called
an $n$-\textbf{chain} in $X$ (of length $m$)   if
\begin{equation}\label{sect2-1}
C_{i_{1} ,\dots ,i_{n} }\cap C_{j_{1} ,\dots ,j_{n} }=\emptyset
\end{equation}
for all $(i_{1} ,\dots ,i_{n} ), (j_{1} ,\dots ,j_{n})\in
\mathbb{N}_{m} ^{n}$ such that $|i_{l} -j_{l} |>1$ for some $l\in
\{1,\dots ,n\}$. Here we use $C_{i_{1} ,\dots ,i_{n}}$ to denote
$C(i_{1} ,\dots ,i_{n})$.

A \textbf{spherical} $(n-1)$-\textbf{chain} in $X$ (of length $m$)
is a function
$$C:\partial \mathbb{N}_{m} ^{n}  \rightarrow \mathcal{P}(X)$$
such that (\ref{sect2-1}) holds for all $(i_{1} ,\dots ,i_{n} ),
(j_{1} ,\dots ,j_{n})\in \partial \mathbb{N}_{m} ^{n}$ such that
$|i_{l} -j_{l} |>1$ for some $l\in \{1,\dots ,n\}$.
\begin{center}
  \includegraphics[height=1.3in,width=1.5in]{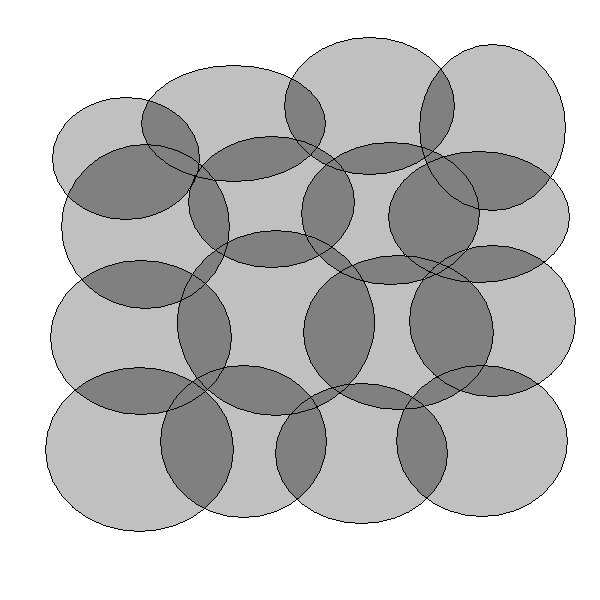}\hspace{25pt}
  \includegraphics[height=1.3in,width=1.5in]{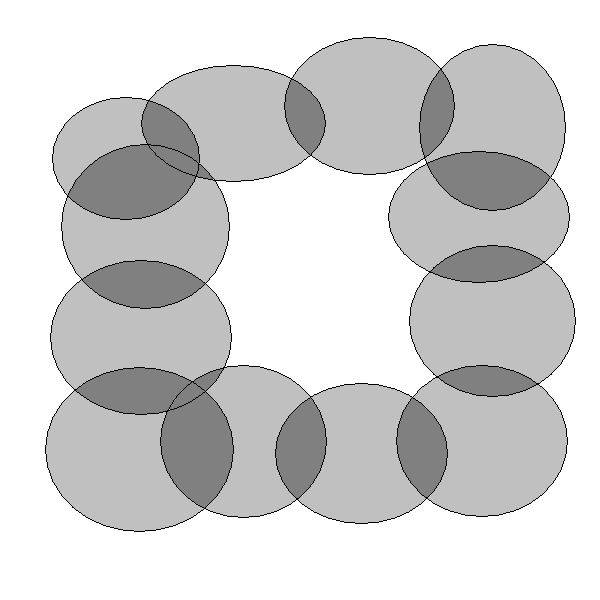}

  \emph{Figure 7.} A 2-chain and a spherical 1-chain
\end{center}

If $C:\mathbb{N}_{m} ^{n} \rightarrow \mathcal{P}(X)$ is a
function, we define its \textbf{boundary} $\partial C$  as the
restriction of $C$ to $\partial \mathbb{N}_{m} ^{n} $. For $i\in
\{1,\dots ,n\}$ and $\rho \in \{0,1\}$ we define $\partial _{i}
^{\rho }C$ as the restriction of $C$ to $\partial _{i}^{\rho
}\mathbb{N}_{m} ^{n} $. Note: if $C$ is an $n$-chain, then
$\partial C$ is a spherical $(n-1)$-chain.

If $C:\partial \mathbb{N}_{m} ^{n} \rightarrow \mathcal{P}(X)$ is
a function and $i\in \{1,\dots ,n\},$ $\rho \in \{0,1\}$, we also
use $\partial _{i} ^{\rho }C$ to denote the restriction of the
function $C$ to $\partial _{i} ^{\rho }\mathbb{N}_{m} ^{n} $.

If $(X,d)$ is a metric space, then we say that an $n-$chain
$C=(C_{i_{1} ,\dots ,i_{n} })_{0\leq i_{1} ,\dots ,i_{n} \leq m}$
in $X$ is \textbf{open} if $C_{i_{1} ,\dots ,i_{n} }$ is an open
set in $(X,d)$ for all $i_{1} ,\dots ,i_{n} \in  \mathbb{N}_{m} $.
We similarly define the notion of a \textbf{compact} $n$-chain in
$(X,d)$ and the notions of open spherical $n$-chain and compact
spherical $n$-chain.

In general, if $A$ is a set and $f:A\rightarrow \mathcal{P}(X)$ a
function, we will denote by $\bigcup f$ the union $\bigcup_{a\in
A}f(a)$ and we will say that $f$ \textbf{covers} $S$, where $S
\subseteq   X$, if $S\subseteq \bigcup f$. If $(X,d)$ is a metric
space and $f(a)$ a nonempty bounded set for each $a\in A$, then we
define $\mesh(f)$ as the number $$\mesh(f)=\max_{a\in
A}\left(\diam f(a)\right).$$

Let $\varepsilon >0$. A (spherical) $n$-chain $C$ in a metric
space $(X,d)$ is said to be a (spherical) $\varepsilon
-n$-\textbf{chain} if $\mesh(C)<\varepsilon $.

 A function $C:A\rightarrow \mathcal{P}(X)$, where $A\subseteq \mathbb{N}_{m} ^{n} $, is said to be
$\varepsilon -$\textbf{proper} if  for all $(i_{1},\dots ,i_{n})$,
 $(j_{1} ,\dots ,j_{n})\in A$ such that $|i_{1} -j_{1} |\leq
1$,\dots ,$|i_{n} -j_{n} |\leq 1$ there exist $x\in C_{i_{1}
,\dots ,i_{n} }$ and $y\in  C_{j_{1} ,\dots ,j_{n} }$ such that
$d(x,y)<\varepsilon $.

The proof of the following lemma is straightforward.
\begin{lem}\label{lema-nova}
Let $(X,d)$ be a metric space, $\varepsilon >0$ and $A\subseteq
\mathbb{N}_{m} ^{n} $. Let $C,D:A\rightarrow \mathcal{P}(X)$ be
such that $D(a)\neq\emptyset $ and $D(a)\subseteq C(a)$ for each
$a\in A$. Suppose $C$ is $\varepsilon -$proper and
$\mesh(C)<\varepsilon $. Then $D$ is $3\varepsilon -$proper. \qed
\end{lem}

\begin{lem}\label{sect2-3}
Let $(X,d)$ be a metric space and let $K$ be a compact (spherical)
$n$-chain of length $m$ in $(X,d)$. Suppose $U_{1} ,\dots ,U_{k} $
are open sets. Then there exists an open (spherical) $n$-chain $C$
of length $m$ in $(X,d)$ such that $K_{a}\subseteq C_{a}$ for all
$a\in \mathbb{N}_{m} ^{n} $ and such that $C_{a}\subseteq U_{i} $
whenever $i\in \{1,\dots ,k \}$ is such that $K_{a}\subseteq U_{i}
$. Moreover, if $\mesh(K)<r $, we can choose $C$ so that
$\mesh(C)<2r $.
\end{lem}
\proof If $S\subseteq X$ and $\varepsilon >0$ let
$$S_{\varepsilon }=\bigcup_{s\in S} B(s,\varepsilon ).$$ This is
clearly an open set. If $S$ is a compact set contained is some
open set $V$, then there exists $\varepsilon
>0$ such that $S_{\varepsilon }\subseteq V$. Furthermore, if $S$
and $T$ are disjoint compact sets, then there exists $\varepsilon
>0$ such that $S_{\varepsilon }\cap T_{\varepsilon }=\emptyset $.
It follows readily from this that there exists $\varepsilon >0 $
such that $C:\mathbb{N}_{m} ^{n} \rightarrow \mathcal{P}(X)$ (or
$C:\partial \mathbb{N}_{m} ^{n} \rightarrow \mathcal{P}(X)$)
defined by $C_{a}=(K_{a})_{\varepsilon }$ is a desired $n$-chain
(spherical $n$-chain). \qed

\begin{prop}\label{sfera-1} Let $n\geq 2$. Suppose
$f:\partial I^{n}\rightarrow S$ is a homeomorphism, where $S$ is a
subspace of a metric space $(X,d)$. Let $U_{i}^{\rho }$, $1\leq
i\leq n$, $\rho \in \{0,1\}$, be open sets in $(X,d)$ such that
$$f(A_{i} ^{\rho })\subseteq U_{i}^{\rho }$$ for all $i\in
\{1,\dots ,n\}$ and $\rho \in \{0,1\}$. Then for each $\varepsilon
>0$ there exists an open spherical $\varepsilon -(n-1)$-chain $C$
in $(X,d)$ which is $\varepsilon -$proper, which covers $S$ and
such that
$$f(A_{i} ^{\rho })\subseteq \bigcup (\partial _{i} ^{\rho
}C)\subseteq U_{i} ^{\rho }$$ for all $i\in \{1,\dots ,n\}$ and
$\rho \in \{0,1\}$.
\end{prop}
\proof For $m\in \mathbb{N}$ let $D^{m} :\mathbb{N}_{m} ^{n}
\rightarrow \mathcal{P}(I^{n} )$ be defined by $$D^{m}_{i_{1}
,\dots ,i_{n} }=\left[\frac{i_{1} }{m+1},\frac{i_{1}
+1}{m+1}\right]\times \dots \times \left[\frac{i_{n}
}{m+1},\frac{i_{n} +1}{m+1}\right].$$ Then $D^{m} $ is a compact
$n-$chain in $I^{n} $ which covers $I^{n} $. Clearly for each
$\varepsilon >0$ there exists $m\in \mathbb{N}$ such that
$\mesh(D^{m} )<\varepsilon $. Note that for all $(i_{1},\dots
,i_{n} ),(j_{1} ,\dots ,j_{n})\in \mathbb{N}_{m} ^{n} $ such that
$|i_{1} -j_{1} |\leq 1$,\dots ,$|i_{n} -j_{n} |\leq 1$ we have
$$D^{m} _{i_{1} ,\dots ,i_{n} }\cap D^{m} _{j_{1} ,\dots ,j_{n}
}\neq\emptyset.$$ We easily conclude from this that for each
$\varepsilon >0$ there exists $m\in \mathbb{N}$ such that
$\mesh(D^{m} )<\varepsilon $ and such that $D^{m} $ is
$\varepsilon -$proper.

The boundary $\partial D^{m} $ is a spherical $(n-1)$-chain in
$I^{n} $ which covers $\partial I^{n} $.

 For $m\in \mathbb{N}$ let
$G^{m} :\partial \mathbb{N}_{m} ^{n} \rightarrow
\mathcal{P}(\partial I^{n} )$ be defined by
$$G^{m} (a)=(\partial D^{m} )(a)\cap \partial I^{n} .$$
Then $G^{m} $ is a compact spherical $(n-1)$-chain in $\partial
I^{n}$ which covers $\partial I^{n} $, moreover
$$A_{i} ^{\rho }\subseteq \bigcup \left(\partial _{i} ^{\rho }G^{m}\right) $$
for all $i\in \{1,\dots ,n\}$ and $\rho \in \{0,1\}$. Note that
these sets need not be equal, however:
\begin{equation}\label{sect2-2}
\mbox{for each }x\in \bigcup \left(\partial _{i} ^{\rho
}G^{m}\right)\mbox{ there exists }y\in A_{i} ^{\rho }\mbox{ such
that }d'(x,y)\leq \frac{1}{m+1},
\end{equation}
where $d'$ is the Euclidean metric on $\mathbb{R}^{n} $. We also
have that for each $\varepsilon
>0$ there exists $m\in \mathbb{N}$ such that $\mesh(G^{m}
)<\varepsilon $ and such that $G^{m} $ is $\varepsilon -$proper.

For $m\in \mathbb{N}$ let $F^{m} :\partial \mathbb{N}_{m} ^{n}
\rightarrow \mathcal{P}(S)$ be defined by
$$F^{m}(a)=f(G^{m}(a)).$$ Then $F^{m} $ is a compact spherical
$(n-1)$-chain in $S$ which covers $S$ and such that
$$f(A_{i} ^{\rho })\subseteq \bigcup \left(\partial _{i} ^{\rho }F^{m}\right) $$
for all $i\in \{1,\dots ,n\}$ and $\rho \in \{0,1\}$.
\begin{center}
  \includegraphics[height=0.9in,width=1in]{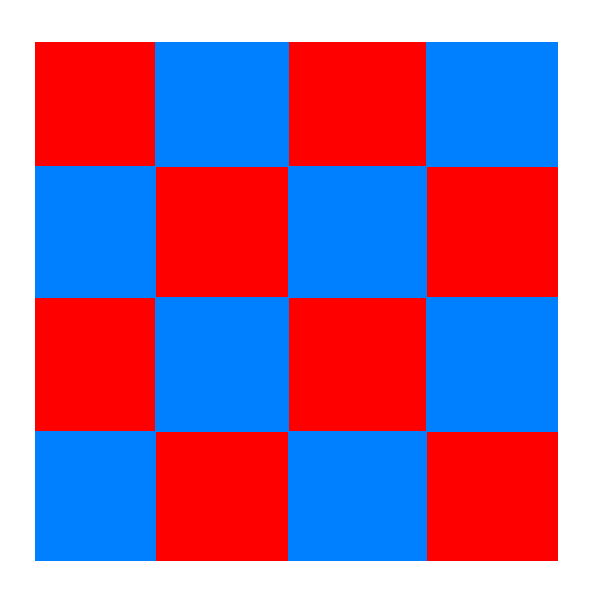}\hspace{33pt}
  \includegraphics[height=0.9in,width=1in]{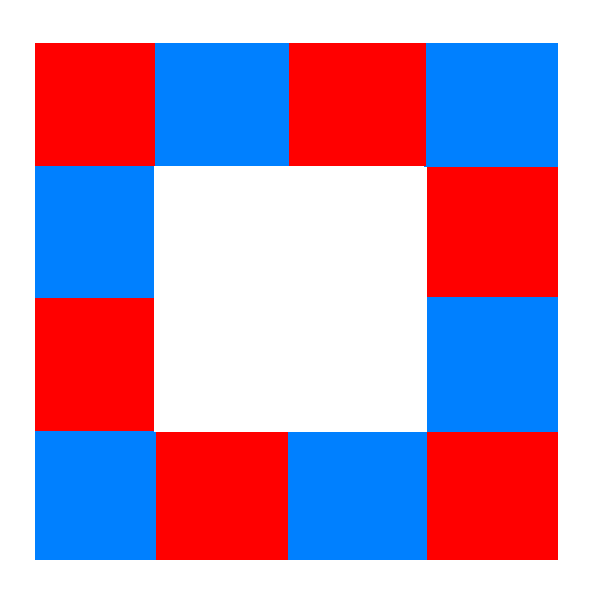}\hspace{33pt}
  \includegraphics[height=0.9in,width=1in]{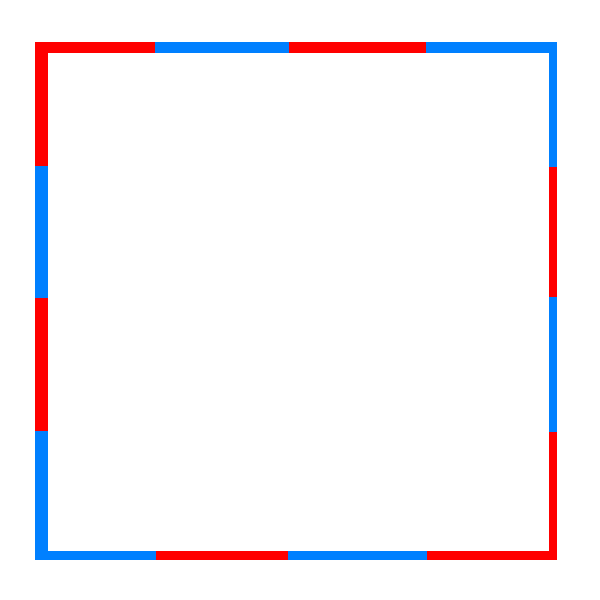}\hspace{33pt}
  \includegraphics[height=0.9in,width=1in]{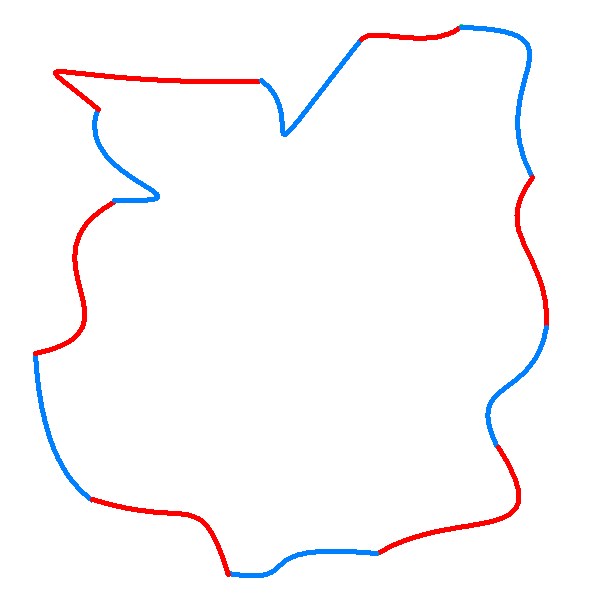}

  \emph{Figure 8.} $D^{3}$, $\partial D^{3}$, $G^{3}$ and $F^{3}$
  (in the case $n=2$)
\end{center}
 The fact
that $f$ is uniformly continuous implies, together with
(\ref{sect2-2}), that for each $\varepsilon
>0$ there exists $m\in \mathbb{N}$ with the property that $F^{m} $ is $\varepsilon -$proper, $\mesh(F^{m} )<\varepsilon
$ and with the property that for all $i\in \{1,\dots ,n\}$, $\rho
\in \{0,1\}$ and $x\in \bigcup \left(\partial _{i} ^{\rho
}F^{m}\right)$ there exists $y\in f(A_{i} ^{\rho })$ such that
$d(x,y)<\varepsilon $.

Let $\varepsilon >0$. Using the fact that the sets $A_{i} ^{\rho
}$ are compact and $U_{i} ^{\rho }$ are open, it is not hard to
conclude now that there exists $m\in \mathbb{N}$ such that $F^{m}
$ is $\varepsilon -$proper, $\mesh(F^{m} )<\frac{\varepsilon}{2} $
and
$$f(A_{i} ^{\rho })\subseteq \bigcup \left(\partial _{i} ^{\rho }F^{m}\right)\subseteq U_{i} ^{\rho } $$
for all $i\in \{1,\dots ,n\}$ and $\rho \in \{0,1\}$. Now we apply
Lemma \ref{sect2-3} to $F$ and the sets $U_{i} ^{\rho }$ and we
get an open spherical $\varepsilon -(n-1)$-chain $C$ in $(X,d)$
which is $\varepsilon -$proper such that
$$f(A_{i}
^{\rho })\subseteq \bigcup (\partial _{i} ^{\rho }C)\subseteq
U_{i} ^{\rho }$$ for all $i\in \{1,\dots ,n\}$ and $\rho \in
\{0,1\}$. \qed

In the same way we prove the following proposition.

\begin{prop}\label{kugla-1} Let $n\geq 1$. Suppose
$f:I^{n}\rightarrow E$ is a homeomorphism, where $E$ is a subspace
of a metric space $(X,d)$. Let $U_{i}^{\rho }$, $1\leq i\leq n$,
$\rho \in \{0,1\}$, be open sets in $(X,d)$ such that
$$f(A_{i} ^{\rho })\subseteq U_{i}^{\rho }$$ for all $i\in
\{1,\dots ,n\}$ and $\rho \in \{0,1\}$. Then for each $\varepsilon
>0$ there exists an open $\varepsilon -n$-chain $C$ in $(X,d)$
which is $\varepsilon -$proper, which covers $E$ and such that
$$f(A_{i} ^{\rho })\subseteq \bigcup (\partial _{i} ^{\rho
}C)\subseteq U_{i} ^{\rho }$$ for all $i\in \{1,\dots ,n\}$ and
$\rho \in \{0,1\}$.
\end{prop}

If $(X,d)$ is a metric space, then for nonempty subsets $S$ and
$T$ of $X$ we denote the number $\inf\{d(x,y)\mid x\in S,y\in T\}$
by $d(S,T)$.

The next proposition provides conditions under which a spherical
$(n-1)$-chain approximates an $(n-1)$-sphere.
\begin{prop} \label{sfera-2}
Let $f:\partial I^{n}\rightarrow S$ be a homeomorphism, where $S$
is a subspace of a metric space $(X,d)$. Let $W_{i}^{\rho }$,
$1\leq i\leq n$, $\rho \in \{0,1\}$, be open sets in $(X,d)$ such
that $W^{0}_{i}\cap W_{i} ^{1} =\emptyset $ for all $i\in
\{1,\dots ,n\}$. Let $\varepsilon >0$ be such that
\begin{equation}\label{sect2-13}
2\varepsilon <d(f(A_{i} ^{0}),f(A_{i} ^{1}))
\end{equation}
 for each $i\in
\{1,\dots ,n\}$. Suppose $C$ is an open spherical $\varepsilon
-(n-1)$-chain  in $(X,d)$ of length $m$ which is
$\varepsilon-$proper, which covers $S$ and suppose that $$f(A_{i}
^{\rho })\subseteq W_{i} ^{\rho },~\bigcup (\partial _{i} ^{\rho
}C)\subseteq W_{i} ^{\rho }$$ for all $i\in \{1,\dots ,n\}$ and
$\rho \in \{0,1\}$. Then for each $x\in \bigcup C$ there exists
$y\in S$ such that $d(x,y)<3\varepsilon $.
\end{prop}
\proof It is enough to prove the following: for each $(p_{1}
,\dots ,p_{n} ) \in \partial \mathbb{N}_{m} ^{n} $ with the
property that $p_{k} \in \{0,m\}$ for exactly one $k\in \{1,\dots
,n\}$ the set $C_{p_{1},\dots ,p_{n}}$ intersects $S$. Namely, if
this holds, then for each $(q_{1} ,\dots ,q_{n} ) \in \partial
\mathbb{N}_{m} ^{n} $ there exists $(p_{1} ,\dots ,p_{n} ) \in
\partial \mathbb{N}_{m} ^{n} $ such that $|q_{1} -p_{1}| \leq 1,$ \dots , $|q_{n} -p_{n}|\leq 1$
and such that $C_{p_{1},\dots ,p_{n}}\cap S\neq\emptyset $. Since
$C$ is an $\varepsilon -(n-1)$-chain and $\varepsilon -$proper, we
now easily get that for each $x\in \bigcup C$ there exists $y\in
S$ such that $d(x,y)<3\varepsilon $.

Suppose the opposite, that there exists  $(p_{1} ,\dots ,p_{n}
)\in
\partial \mathbb{N}_{m} ^{n}$ such that $p_{k} \in \{0,m\}$ for exactly one $k\in
\{1,\dots ,n\}$ and such that $C_{p_{1},\dots ,p_{n}}\cap
S=\emptyset$.  We may assume  $p_{n} =m$ (all other cases can be
reduced to this one if we modify $C$ and $f$ by interchange of
appropriate coordinates). It follows $0<p_{1}<m$, \dots ,
$0<p_{n-1}<m$.

For $i\in \{1,\dots ,n-1\}$ we define the set $U_{i} $ as the
union of all sets of the following form:
\begin{equation}\label{sect2-10}
C_{j_{1} ,\dots ,j_{i-1},l,j_{i+1},\dots ,j_{n-1},m},\mbox{ where
}l<p_{i} ;
\end{equation}
\begin{equation}\label{sect2-11}
C_{j_{1} ,\dots ,j_{i-1},0,j_{i+1},\dots ,j_{n} };
\end{equation}
\begin{equation}\label{sect2-12}
C_{j_{1} ,\dots ,j_{n-1},0},\mbox{ where this set is such that it
intersects }f(A_{i} ^{0}).
\end{equation}
Furthermore, let $V_{i} $ be the union of all sets of the
following form:
\begin{equation}\label{sect2-14}
C_{j_{1} ,\dots ,j_{i-1},l,j_{i+1},\dots ,j_{n-1},m},\mbox{ where
}l>p_{i} ;
\end{equation}
\begin{equation}\label{sect2-15}
C_{j_{1} ,\dots ,j_{i-1},m,j_{i+1},\dots ,j_{n} };
\end{equation}
\begin{equation}\label{sect2-17}
C_{j_{1} ,\dots ,j_{n-1},0},\mbox{ where this set is such that it
intersects }f(A_{i} ^{1}).
\end{equation}

Let $i\in \{1,\dots ,n-1\}$. The sets $U_{i} $ and $V_{i} $ are
open and it is straightforward to check that they are disjoint.
(Figures 9.\ and 10.\ show $C$ in case $n=3$ and $m=5$; the red
set is $C_{p_{1} ,p_{2} ,p_{3} }$, in this case $C_{2,2,5}$, the
blue sets in Figure 9.\ are $U_{1} $ and $V_{1} $, the blue sets
in Figure 10.\ are $U_{2} $ and $V_{2} $. For example, note that
in Figure 11. the black set
 is $C_{3,0,2}$, the red set is $C_{5,0,1}$ and the blue set is $C_{5,0,5}$.)
\begin{center}
 \hspace{30pt} \includegraphics[height=1.7in,width=2in]{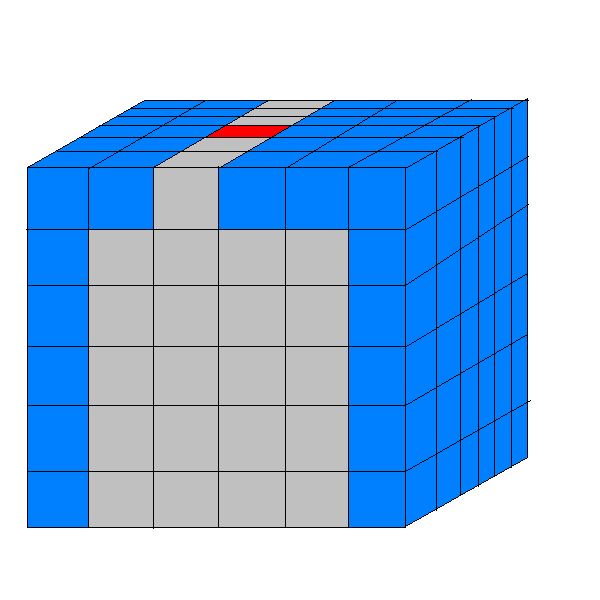}\hspace{30pt}
  \includegraphics[height=1.7in,width=2in]{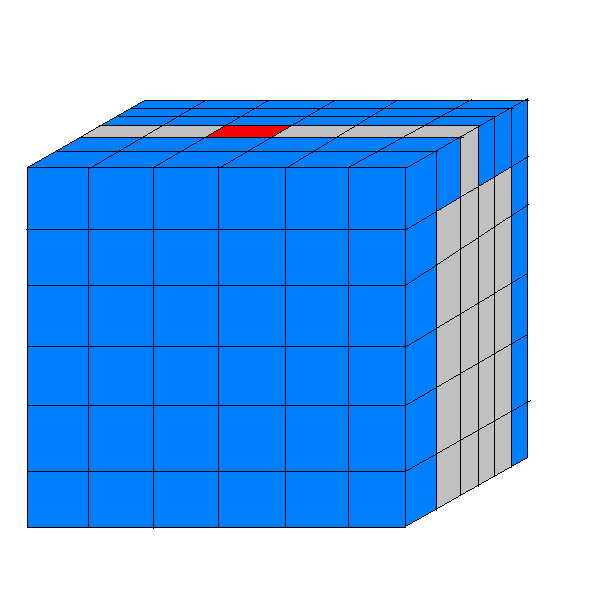}

  \emph{Figure 9.}  \hspace{120pt} \emph{Figure
  10.}
\end{center}

\begin{center}
 \includegraphics[height=1.7in,width=2in]{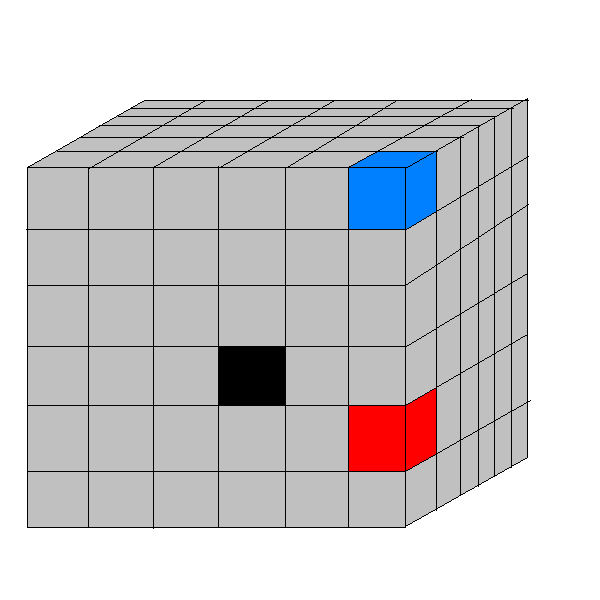}

  \emph{Figure 11.}
\end{center}

We also have
\begin{equation}\label{sect2-18}
U_{i} \cap f(A_{i} ^{1}\cap A_{n} ^{0})=\emptyset.
\end{equation}
Otherwise, the set $f(A_{i} ^{1}\cap A_{n} ^{0})=f(A_{i} ^{1})\cap
f(A_{n} ^{0})$ would intersect one of the sets in
(\ref{sect2-10}), (\ref{sect2-11}) or (\ref{sect2-12}). The sets
in (\ref{sect2-10}) are contained in $W_{n} ^{1}$ which is
disjoint with $f(A_{n} ^{0})$. The sets in (\ref{sect2-11}) are
contained in $W_{i} ^{0}$ which is disjoint with $f(A_{i} ^{1})$.
Finally, $f(A_{i} ^{1})$ cannot intersect a set in
(\ref{sect2-12}) since (\ref{sect2-13}) holds. In the same way we
get
\begin{equation}\label{sect2-19}
V_{i}\cap f(A_{i} ^{0}\cap A_{n} ^{0})=\emptyset .
\end{equation}

Let
$$\Omega =U_{1} \cup \dots \cup
U_{n-1}\cup V_{1} \cup \dots \cup V_{n-1}.$$ Let $i\in \{1,\dots
,n\}$ and $\rho \in \{0,1\}$ such that $(i,\rho )\neq (n,0)$. We
claim that
\begin{equation}\label{sect2-16}
f(A_{i} ^{\rho })\subseteq \Omega .
\end{equation}
Suppose that there exists $x\in f(A_{i}^{\rho })$ such that
$x\notin \Omega $. Since $C$ covers $S$, there exists $(j_{1}
,\dots ,j_{n}) \in
\partial \mathbb{N}_{m}^{n}  $ such that
$$x\in C_{j_{1} ,\dots ,j_{n} }.$$ We have $(j_{1} ,\dots ,j_{n}
)\neq (p_{1} ,\dots ,p_{n} )$ since $C_{p_{1} ,\dots ,p_{n} }\cap
S=\emptyset $. So, if $j_{n} =m$, then $C_{j_{1} ,\dots ,j_{n}}$
must be one of the sets in (\ref{sect2-10}) or (\ref{sect2-14}).
But this is impossible since $x\notin \Omega $. So $j_{n} <m$.
Now, if $j_{n} >0$, then $C_{j_{1} ,\dots ,j_{n}}$ is one of the
sets in (\ref{sect2-11}) or (\ref{sect2-15}), impossible.
Therefore $j_{n} =0$.

We have $C_{j_{1} ,\dots ,j_{n}}\cap f(A_{i} ^{\rho
})\neq\emptyset $ and this also yields to contradiction. Namely,
if $i<n$, then $C_{j_{1} ,\dots ,j_{n}}$ is one of the sets in
(\ref{sect2-12}) or (\ref{sect2-17}). And if $i=n$, then $\rho =1$
and $$f(A_{n}^{1})\subseteq W_{n} ^{1},~C_{j_{1} ,\dots
,j_{n}}\subseteq \bigcup(\partial _{n} ^{0}C)\subseteq W_{n}
^{0},$$ which is impossible since $W_{n} ^{0}\cap W_{n}
^{1}=\emptyset $. Hence (\ref{sect2-16}) holds.

Let $$E=A_{1} ^{0}\cup \dots \cup A_{n-1}^{0}\cup A_{1} ^{1}\cup
\dots \cup A_{n} ^{1}.$$ For each $i\in \{1,\dots ,n-1\}$ the sets
$f^{-1} (U_{i} )$ and $f^{-1} (V_{i} )$ are open in $\partial
I^{n} $, they are disjoint, by (\ref{sect2-18}) and
(\ref{sect2-19})
$$f^{-1} (U_{i} )\cap (A_{i} ^{1}\cap A_{n} ^{0})=\emptyset ,~f^{-1} (V_{i} )\cap (A_{i} ^{0}\cap A_{n} ^{0})=\emptyset$$
and by (\ref{sect2-16}) $$E\subseteq f^{-1} (U_{1} )\cup \dots\cup
f^{-1} (U_{n-1} )\cup f^{-1} (V_{1} )\cup \dots\cup f^{-1}
(V_{n-1} ).$$ This is impossible by Corollary \ref{kor-glavni}.
\qed

The next proposition provides conditions under which an $n$-chain
approximates an $n$-cell.
\begin{prop}\label{kugla-2}
Let $f:I^{n}\rightarrow E$ be a homeomorphism, where $E$ is a
subspace of a metric space $(X,d)$.
Suppose $C$ is an open  $\varepsilon -n$-chain  in $(X,d)$ of
length $m$ which is $\varepsilon -$proper, which covers $E$ and
such that $\partial C$ covers $f(\partial I^{n} )$ and suppose
that $W_{i}^{\rho }$, $1\leq i\leq n$, $\rho \in \{0,1\}$, are
open sets in $(X,d)$ such that
$$2\varepsilon <d(W_{i} ^{0},W_{i} ^{1}),$$ $$f(A_{i} ^{\rho
})\subseteq W_{i} ^{\rho }\mbox{ and }\bigcup (\partial _{i}
^{\rho }C)\subseteq W_{i} ^{\rho }$$ for all $i\in \{1,\dots ,n\}$
and $\rho \in \{0,1\}$. Then for each $x\in \bigcup C$ there
exists $y\in E$ such that $d(x,y)<7\varepsilon $.
\end{prop}
\proof It is enough to prove that for each $(p_{1} ,\dots ,p_{n} )
\in \mathbb{N}_{m} ^{n} $ with the property that $1<p_{k} <m-1$
for each $k\in \{1,\dots ,n\}$ the set $C_{p_{1},\dots ,p_{n}}$
intersects $E$. Why is it enough to prove this? Suppose that this
fact holds. Assume that $m\geq 4$. Let $x\in \bigcup C$. Then
$x\in C_{q_{1},\dots ,q_{n}}$ for some $(q_{1} ,\dots ,q_{n}
)\in\mathbb{N}_{m} ^{n}$. For $i\in \{1,\dots ,n\}$ let $$q_{i}
'=\left\{\begin{tabular}{l}
                $1$ if $q_{i} =0,$\\
                $m-1$ if $q_{i} =m,$\\
                $q_{i} $ otherwise.
              \end{tabular} \right. $$
Then $|q_{1} -q_{1} '|\leq 1$, \dots, $|q_{n} -q_{n} '|\leq 1$ and
$1\leq q_{i} '\leq m-1$ for each $i\in \{1,\dots ,n\}$. Since $C$
is $\varepsilon -$proper, there exist $x'\in C_{q_{1} ,\dots
,q_{n} }$ and $x''\in C_{q_{1} ',\dots ,q_{n} '}$ such that
$d(x',x'')<\varepsilon $. Now, let the numbers $p_{1} ,\dots
,p_{n} $ be defined by $$p_{i} =\left\{\begin{tabular}{l}
                $2$ if $q_{i}' =1,$\\
                $m-2$ if $q_{i}' =m-1,$\\
                $q_{i}' $ otherwise,
              \end{tabular} \right. $$ $i\in \{1,\dots ,n\}$. We
have $|q_{1}' -p_{1} |\leq 1$, \dots, $|q_{n}' -p_{n} |\leq 1|$
and therefore there exist $y''\in C_{q_{1} ',\dots ,q_{n} '}$ and
$y'\in C_{p_{1} ,\dots ,p_{n} }$ such that $d(y'',y')<\varepsilon
$. Clearly $1<p_{i} <m-1$ for each $i\in \{1,\dots ,n\}$ and
therefore there exists $y\in C_{p_{1} ,\dots ,p_{n} }\cap E$.
Using the fact that the diameters of the sets $C_{q_{1} ,\dots
,q_{n} }$, $C_{q_{1}' ,\dots ,q_{n}' }$ and $C_{p_{1} ,\dots
,p_{n} }$ are less than $\varepsilon $, we obtain $$d(x,y)\leq
d(x,x')+d(x',x'')+d(x'',y'')+d(y'',y')+d(y',y)<5\varepsilon .$$

If $m\leq 3$, then for all $(a_{1} ,\dots ,a_{n} ), (b_{1} ,\dots
,b_{n} )\in \mathbb{N}_{m} ^{n}$ we have, for each $i\in \{1,\dots
,n\}$, that $a_{i} ,b_{i} \in \{0,\dots ,m\}\subseteq \{0,1,2,3\}$
and therefore there exist $c_{i} ,d_{i} \in \{0,\dots ,m\}$ such
that $|a_{i} -c_{i} |\leq 1$, $|c_{i} -d_{i} |\leq 1$, $|d_{i}
-b_{i} |\leq 1$ which, together with the fact that $C$ is
$\varepsilon -$proper, implies that there exist $x\in C_{a_{1}
,\dots ,a_{n} }$, $x',y'\in C_{c_{1} ,\dots ,c_{n} }$, $y'',z''\in
C_{d_{1} ,\dots ,d_{n} }$ and $z\in C_{b_{1} ,\dots ,b_{n} }$ such
that $d(x,x')<\varepsilon $, $d(y',y'')<\varepsilon $ and
$d(z'',z)<\varepsilon $. It follows $d(x,z)<5\varepsilon $. This
proves that $d(C_{a},C_{b})<5\varepsilon $ for all $a,b\in
\mathbb{N}_{m} ^{n}$. Since $E$ is nonempty and contained in
$\bigcup C$, there exists $b\in \mathbb{N}_{m} ^{n}$ such that
$C_{b}\cap E\neq\emptyset $. It follows $d(C_{a},E)<6\varepsilon $
for each $a\in \mathbb{N}_{m} ^{n}$ and therefore for each $x\in
\bigcup C$ we have $d(x,E)<7\varepsilon $.

So, let  $(p_{1} ,\dots ,p_{n} ) \in \mathbb{N}_{m} ^{n} $ be such
that $1<p_{k} <m-1$ for each $k\in \{1,\dots ,n\}$. We want to
prove that $C_{p_{1},\dots ,p_{n}}\cap E\neq\emptyset $. Suppose
$C_{p_{1},\dots ,p_{n}}\cap E=\emptyset $.

For $i\in \{1,\dots ,n\}$ let $U_{i} $ be the union of all sets
$C_{j_{1} ,\dots ,j_{n} }$ such that
\begin{equation}\label{sect2-20-1}
j_{i} <p_{i}\mbox{ and }1< j_{k} < m-1\mbox{ for all }k\in
\{1,\dots ,n\}
\end{equation}
 or
\begin{equation}\label{sect2-20-2}
j_{i} \in \{0,1\}.
\end{equation}
Let $V_{i} $ be the union of all sets $C_{j_{1} ,\dots ,j_{n} }$
such that
$$j_{i} >p_{i}\mbox{ and }1< j_{k} < m-1\mbox{ for all }k\in
\{1,\dots ,n\}$$ or $$j_{i} \in \{m-1,m\}.$$ For each $i\in
\{1,\dots ,n\}$ the sets $U_{i} $ and $V_{i} $ are open and
disjoint. Note that every $C_{j_{1} ,\dots ,j_{n} }$, where
$(j_{1} ,\dots ,j_{n} )\neq (p_{1} ,\dots ,p_{n} )$, is contained
in some $U_{i} $ or $V_{i} $. Therefore
\begin{equation}\label{sect2-20}
E\subseteq U_{1} \cup \dots \cup U_{n} \cup V_{1} \cup \dots \cup
V_{n} .
\end{equation}
 Let $i\in \{1,\dots ,n\}$. We prove now that
\begin{equation}\label{sect2-21}
U_{i} \cap f(A_{i} ^{1})=\emptyset .
\end{equation}
Suppose the opposite, that $U_{i} \cap f(A_{i}
^{1})\neq\emptyset$. It follows from the definition of $U_{i} $
that there exist $j_{1} ,\dots ,j_{n} \in \mathbb{N}_{m} $  such
that (\ref{sect2-20-1}) or (\ref{sect2-20-2}) hold and such that
$C_{j_{1} ,\dots ,j_{n} }\cap f(A_{i} ^{1})\neq\emptyset$.
However, if (\ref{sect2-20-1}) holds, then $1< j_{k} < m-1$ for
all $k\in \{1,\dots ,n\}$ which implies that $C_{j_{1} ,\dots
,j_{n} }$ is disjoint with $\bigcup (\partial C)$. But we have the
assumption that $\partial C$ covers $f(\partial I^{n} )$ and this
implies that $C_{j_{1} ,\dots ,j_{n} }$ is disjoint with
$f(\partial I^{n} )$ which is impossible. Therefore,
(\ref{sect2-20-1}) does not hold which means that
(\ref{sect2-20-2}) holds. Hence we have
$$C_{j_{1} ,\dots ,j_{n} }\cap f(A_{i} ^{1})\neq \emptyset\mbox{
and }j_{i} \in \{0,1\}.$$ This, together with
$\bigcup(\partial_{i} ^{0} C)\subseteq W_{i} ^{0}$ and $f(A_{i}
^{1})\subseteq W_{i} ^{1}$, implies $d(W_{i} ^{0},W_{i}
^{1})<2\varepsilon $ (namely, if $j_{i} =0$, then $W_{i} ^{0}\cap
W_{i} ^{1}\neq\emptyset $, and if $j_{i} =1$, then $d(C_{j_{1}
,\dots ,j_{n} },W_{i} ^{0})<\varepsilon $ since $C$ is
$\varepsilon -$proper and this implies $d(W_{i} ^{0},W_{i}
^{1})<2\varepsilon $). A contradiction. Hence (\ref{sect2-21})
holds. In the same way we get
\begin{equation}\label{sect2-22}
V_{i} \cap f(A_{i} ^{0})=\emptyset .
\end{equation}

For each $i\in \{1,\dots ,n\}$ the sets $f^{-1} (U_{i} )$ and
$f^{-1} (V_{i} )$ are open in $I^{n} $ and disjoint. By
(\ref{sect2-20})
$$I^{n} =f^{-1}(U_{1} )\cup \dots \cup f^{-1} (U_{n} )\cup f^{-1}
(V_{1} )\cup \dots \cup f^{-1} (V_{n} ),$$ and by (\ref{sect2-21})
and (\ref{sect2-22}) $$f^{-1} (U_{i}) \cap A_{i}
^{1}=\emptyset,~f^{-1} (V_{i}) \cap A_{i} ^{0}=\emptyset.$$ This
is impossible by Corollary \ref{glavni}. \qed


\section{Computability of co-c.e.\ spheres and cells}\label{sect-comp}

Let $n\geq 1$. A \textbf{finite} $n-$\textbf{sequence} in
$\mathbb{N}$ is any function of the form $$\{0,\dots ,m\}^{n}
\rightarrow \mathbb{N}.$$ Recall that any finite sequence $i_{0}
,\dots ,i_{m} $ in $\mathbb{N}$ is of the form $(j)_{0} ,\dots,
(j)_{\overline{j}}$ for some $j\in \mathbb{N}$. Let
$f:\mathbb{N}^{n} \rightarrow \mathbb{N}$ be some computable
injection and let $\tau $ and $\tau '$ be the functions from the
section Preliminaries. We define
$\Sigma:\mathbb{N}^{n+1}\rightarrow \mathbb{N}$ by
$$\Sigma(i,j_{1} ,\dots ,j_{n} )=(\tau (i))_{f(j_{1} ,\dots ,j_{n}
)}.$$ Then for any finite $n-$sequence $a$ in $\mathbb{N}$ there
exists $i\in \mathbb{N}$ such that $a$ equals the function
$$\{0,\dots ,\tau '(i)\}^{n} \rightarrow \mathbb{N},$$ $$(j_{1} ,\dots
,j_{n} )\mapsto \Sigma(i,j_{1} ,\dots ,j_{n} ).$$ We will use the
following notation: $\widehat{i}$ instead of $\tau '(i)$ and, for
$n\geq 2$, $(i)_{j_{1} ,\dots ,j_{n} }$ instead of $\Sigma(i,j_{1}
,\dots ,j_{n} )$.

Let $(X,d,\alpha )$ be a computable metric space. For $l\in
\mathbb{N}$ let $\mathcal{H}_{l}$ be the finite $n-$sequence of
sets in $X$ defined by
$$\mathcal{H}_{l}=\left(J_{(l)_{j_{1} ,\dots ,j_{n} }}\right)_{0\leq
j_{1} ,\dots ,j_{n} \leq \widehat{l}}$$ (i.e$.$ $\mathcal{H}_{l}$
is the function $\{0,\dots ,\widehat{l}\}^{n} \rightarrow
\mathcal{P}(X)$ which maps $(j_{1} ,\dots ,j_{n} )$ to
$J_{(l)_{j_{1} ,\dots ,j_{n} }}$).

For $l\in \mathbb{N}$ let $\widehat{\mathcal{H}}_{l}$ be defined
by
$$\widehat{\mathcal{H}}_{l}=\left(\widehat{J}_{(l)_{j_{1} ,\dots ,j_{n} }}\right)_{0\leq
j_{1} ,\dots ,j_{n} \leq \widehat{l}}.$$

In Euclidean space $\mathbb{R}^{n}$ we can effectively calculate
the diameter of the finite union of rational balls. However, in a
general computable metric space the function
$\mathbb{N}\rightarrow \mathbb{R}$, $j\mapsto \diam(J_{j} )$, need
not be computable. For that reason we are going to use the notion
of the formal diameter. Let $(X,d)$ be a metric space and
$x_{0},\dots ,x_{k} \in X,$ $r_{0} ,\dots ,r_{k} \in
\mathbb{R}_{+}.$ The \textbf{formal diameter} associated to the
finite sequence $(x_{0},r_{0} ),\dots ,(x_{k},r_{k} ) $ is the
number $D\in \mathbb{R}$ defined by
$$D=\max_{0\leq v,w\leq k }d(x_{v} ,x_{w} )+2\max_{0\leq v\leq
k}r_{v}.$$

Let $(X,d,\alpha )$ be a computable metric space. We define the
function $\fdiam:\mathbb{N}\rightarrow \mathbb{R}$ in the
following way. For $j\in \mathbb{N}$ the number $\fdiam(j)$ is the
formal diameter associated to the finite sequence
$$\left(\alpha _{\tau ((j)_{0})} , q_{\tau
'((j)_{0})}\right),\dots ,\left(\alpha_{ \tau ((j)_{\overline{j}})
}, q_{\tau' ((j)_{\overline{j}})}\right).$$ We have the following
proposition (for the proof see \cite{zi:mib}).
\begin{prop} \label{fdiam}
Let $(X,d,\alpha )$ be a computable metric space.
\begin{enumerate}[\em(1)]
\item For all $j\in \mathbb{N}$, $\diam(\widehat{J}_{j} )\leq
\fdiam(j)$. \item $\fdiam:\mathbb{N}\rightarrow \mathbb{R}$ is a
computable function. \item Let  $S$ be a compact subset of
$(X,d),$ $r\in \mathbb{R}_{+}$ and $C_{0} ,\dots ,C_{m} $ a finite
sequence of open sets which covers $S$ and such that $\diam(C_{i}
)<r$ for each $i\in \{0,\dots ,m\}$.  Then there exist $j_{0}
,\dots ,j_{m} \in \mathbb{N}$ such that the finite sequence of
sets $J_{j_{0} } ,\dots ,J_{j_{m}} $ covers $S$,
$\widehat{J_{j_{i}} }\subseteq C_{i} $ and $\fdiam(j_{i} )<r$ for
each $i\in \{0,\dots ,m\}.$
\end{enumerate}

\end{prop}

\noindent Let the function $\fmesh:\mathbb{N}\rightarrow \mathbb{R}$ be
defined by
$$\fmesh(l)=\max_{0\leq j_{1} ,\dots ,j_{n} \leq \widehat{l}}\fdiam((l)_{j_{1} ,\dots ,j_{n} }).$$
It is immediate from Proposition \ref{fdiam} and Proposition
\ref{NuR} that $\fmesh$ is a computable function.

\begin{prop} \label{prop-proper} Let $(X,d,\alpha )$ be a computable metric space.
The sets $$\Omega =\{(l,k)\in \mathbb{N}^{2}\mid
\mathcal{H}_{l}\mbox{ is }2^{-k}-\mbox{proper}\}$$ and
$$\Omega' =\{(l,k)\in \mathbb{N}^{2}\mid
\partial  \mathcal{H}_{l}\mbox{ is }2^{-k}-\mbox{proper}\}$$ are c.e.
\end{prop}
\proof Let $\Phi :\mathbb{N}^{2}\rightarrow
\mathcal{P}(\mathbb{N}^{2n+2})$ be defined in the following way.
For $l,k\in \mathbb{N}$ let $\Phi (l,k)$ be the set of all
$$(l,k,i_{1} ,..,i_{n} ,j_{1} ,\dots ,j_{n})$$ such that $ i_{1}
,..,i_{n} ,j_{1} ,\dots ,j_{n}\in \mathbb{N}_{\widehat{l}}$ and
$|i_{1} -j_{1} |\leq 1$, \dots, $|i_{n} -j_{n} |\leq 1$. Then
$\Phi $ is c.c.b. On the other hand, let $S$ be the set of all
$(l,k,i_{1} ,\dots ,i_{n} ,j_{1} ,\dots ,j_{n} )$ for which there
exists $x\in J_{(l)_{i_{1} ,\dots ,i_{n}} }$ and $y\in
J_{(l)_{j_{1} ,\dots ,j_{n}} }$ such that $d(x,y)<2^{-k}$. This is
equivalent to the fact that there exist $p,q\in \mathbb{N}$ such
that
\begin{equation}\label{sect3-3}
\alpha _{p}\in J_{(l)_{i_{1} ,\dots ,i_{n}} },~\alpha _{q}\in
J_{(l)_{j_{1} ,\dots ,j_{n}} }\mbox{ and }d(\alpha _{p},\alpha
_{q})<2^{-k}
\end{equation}
The set $T$ of all $(l,k,i_{1} ,\dots ,i_{n} ,j_{1} ,\dots
,j_{n},p,q )$ such that (\ref{sect3-3}) holds is c.e.\ by
Corollary \ref{kJj} and Proposition \ref{NuR}. Therefore $S$ is
c.e. Since
$$\Omega =\{(l,k)\mid \Phi (l,k)\subseteq S\}$$
we have that $\Omega $ is c.e.\ by Proposition \ref{p1}. We
similarly get that $\Omega '$ is c.e. \qed

\begin{lem}\label{l-zeta}
Let $(X,d,\alpha )$ be a computable metric space. There exists a
computable function $\zeta  :\mathbb{N}\rightarrow \mathbb{N}$
such that $J_{\zeta(l)}=\bigcup\mathcal{H}_{l}$ for each $l\in
\mathbb{N}$. There exists a computable function $\zeta
':\mathbb{N}\rightarrow \mathbb{N}$ such that
$J_{\zeta'(l)}=\bigcup(\partial \mathcal{H}_{l})$ for each $l\in
\mathbb{N}$. Furthermore, for all $i\in \{1,\dots ,m\}$ and $\rho
\in \{0,1\}$ there exists a computable function $\zeta
'':\mathbb{N}\rightarrow \mathbb{N}$ such that
$J_{\zeta''(l)}=\bigcup(\partial_{i} ^{\rho } \mathcal{H}_{l})$
for each $l\in \mathbb{N}$. Similar statements hold for
$\widehat{J_{j} }$ and $\widehat{\mathcal{H}}_{l}$, $\partial
\widehat{\mathcal{H}}_{l}$, $\partial _{i} ^{\rho
}\widehat{\mathcal{H}}_{l}$.
\end{lem}
\proof It is enough to prove the following: if $\Phi
:\mathbb{N}\rightarrow \mathcal{P}(\mathbb{N}^{n} )$ and $\Psi
:\mathbb{N}^{n} \rightarrow \mathcal{P}(\mathbb{N})$ are c.c.b.\
functions such that $\Phi (l)\neq\emptyset $ and $\Psi
(a)\neq\emptyset $ for all $l\in \mathbb{N}$ and $a\in
\mathbb{N}^{n} $, then there exists a computable function $\zeta
:\mathbb{N}\rightarrow \mathbb{N}$  such that
\begin{equation}\label{sect3-20}
J_{\zeta (l)}=\bigcup_{a\in \Phi (l)}\bigcup_{i\in \Psi (a)}I_{i}
.
\end{equation}
 However, if $\Phi $ and $\Psi $ are such functions, by
Proposition \ref{p1}  there exists a c.c.b.\ function $\Lambda
:\mathbb{N}\rightarrow \mathcal{P}(\mathbb{N})$ such that
$$\bigcup_{i\in \Lambda (l)}I_{i} =\bigcup_{a\in \Phi
(l)}\bigcup_{i\in \Psi (a)}I_{i} .$$ For each $l\in \mathbb{N}$
there exists $j\in \mathbb{N}$ such that $\Lambda (l)=[j]$ (recall
definition (\ref{p2-eq})). Since the set $S=\{(l,j)\mid \Lambda
(l)=[j]\}$ is computable (Proposition \ref{p1}) and for each $l\in
\mathbb{N}$ there exists $j\in \mathbb{N}$ such that $(l,j)\in S$,
there exists a computable function $\zeta :\mathbb{N}\rightarrow
\mathbb{N}$ such that $(l,\zeta (l))\in S$ for each $l\in
\mathbb{N}$. It follows (\ref{sect3-20}). \qed

The proof of the following proposition can be found in
\cite{zi:mib}.

\begin{prop} \label{pr-S-J}
Let $(X,d,\alpha )$ be a computable metric space which has the
effective covering property and compact closed balls.
\begin{enumerate}[\em(1)]
\item The set $\{(i,j)\in \mathbb{N}^{2}\mid
\widehat{J}_{i}\subseteq J_{j}\}$ is c.e.

\item Let $S$ be a co-c.e$.$ set in $(X,d,\alpha )$ which is
compact. Then the set $\{j\in \mathbb{N}\mid S\subseteq J_{j} \}$
is c.e.\qed
\end{enumerate}
\end{prop}

\begin{cor} \label{kor-pokrivanje}
Let $(X,d,\alpha )$ be a computable metric space which has the
effective covering property and compact closed balls and let $S$
be a co-c.e$.$ set in $(X,d,\alpha )$ which is compact. Then the
sets $$\{l\in \mathbb{N}\mid \mathcal{H}_{l} \mbox{ covers
}S\}\mbox{ and }\{l\in \mathbb{N}\mid \partial \mathcal{H}_{l}
\mbox{ covers }S\}$$ are c.e.
\end{cor}
\proof This follows from Lemma \ref{l-zeta} and Proposition
\ref{pr-S-J}. \qed

The following proposition can be proved in the same way as
Proposition 32 in \cite{zi:mib}.
\begin{prop} \label{prop-chain-sphchain}
Let $(X,d,\alpha )$ be a computable metric space which has the
effective covering property and compact closed balls. The sets
$\Omega =\{l\in \mathbb{N}\mid \widehat{\mathcal{H}}_{l} $ is an
$n$-chain$\}$ and $\Omega '=\{l\in \mathbb{N}\mid
 \partial \widehat{\mathcal{H}}_{l} $ is a spherical $(n-1)$-chain$\}$ are
computably enumerable. \qed
\end{prop}

The following lemma can be proved similarly as Lemma 14 in
\cite{zi:mib}.
\begin{lem} \label{aproks}
Let $(X,d,\alpha )$ be a computable metric space. Let $S$ be a
compact set in this space such that that there exists a computable
function $f:\mathbb{N}\rightarrow \mathbb{N}$ with the property
that for each $k\in \mathbb{N}$ the following holds: $$S\subseteq
J_{f(k)}\mbox{ and for each }x\in J_{f(k)}\mbox{ there exists
}y\in S\mbox{ such that }d(x,y)<2^{-k}.$$ Then $S$ is computable.
\qed
\end{lem}

\begin{thm}\label{thm-1}
Let $(X,d,\alpha )$ be a computable metric space which is locally
computable. Let $S$ be an $(n-1)-$sphere in $(X,d)$ and suppose
$S$ is co-c.e. in $(X,d,\alpha )$. Then $S$ is computable.
\end{thm}
\proof As we have seen, we may assume that $(X,d,\alpha )$ has
compact closed balls and the effective covering property. Let
$f:\partial I^{n} \rightarrow S$ be a homeomorphism. Choose sets
$W_{i}^{\rho }$, $1\leq i\leq n$, $\rho \in \{0,1\}$, so that each
of these sets is a finite union of rational balls (i.e$.$ of the
form $J_{j} $) and so that $$W^{0}_{i}\cap W_{i} ^{1}
=\emptyset\mbox{ and }f(A_{i} ^{\rho })\subseteq W_{i} ^{\rho }$$
for all $i\in \{1,\dots ,n\}$ and $\rho \in \{0,1\}$.

Let $k_{0} \in \mathbb{N}$ be such that
$$2\cdot 2^{-k_{0} }  <d(f(A_{i} ^{0}),f(A_{i} ^{1}))$$
 for each $i\in
\{1,\dots ,n\}$.

By Proposition \ref{sfera-1} for each $\varepsilon >0$ there
exists an open spherical $\varepsilon  -(n-1)$-chain $C$ in
$(X,d)$ which is $\varepsilon  -$proper, which covers $S$ and such
that
$$ f(A_{i} ^{\rho })\subseteq\bigcup (\partial _{i} ^{\rho
}C)\subseteq W_{i} ^{\rho }$$ for all $i\in \{1,\dots ,n\}$ and
$\rho \in \{0,1\}$.

From this, Lemma \ref{lema-nova} and  Proposition \ref{fdiam} we
conclude that for each $k\in \mathbb{N}$ there exists $l\in
\mathbb{N}$ with the following properties:
\begin{equation}\label{tm1}
\partial \widehat{\mathcal{H}}_{l}\mbox{ is a spherical
}(n-1)\mbox{-chain},
\end{equation}
\begin{equation}\label{tm2}
\partial \mathcal{H}_{l}\mbox{ covers }S,
\end{equation}
\begin{equation}\label{tm3}
\fmesh(l)<2^{-(k+k_{0} )},
\end{equation}
\begin{equation}\label{tm4}
\partial \mathcal{H}_{l}\mbox{ is }2^{-(k+k_{0} )}-\mbox{ proper}
\end{equation}
and
\begin{equation}\label{tm5}
\bigcup \left(\partial _{i} ^{\rho
}\widehat{\mathcal{H}}_{l}\right)\subseteq W_{i} ^{\rho }
\end{equation}
for all $i\in \{1,\dots ,n\}$ and $\rho \in \{0,1\}$.

Let $\Omega $ be the set of all $(k,l)$ such that (\ref{tm1}),
(\ref{tm2}), (\ref{tm3}), (\ref{tm4}) and (\ref{tm5}) hold. Then
$\Omega $ is c.e., which follows from  Proposition
\ref{prop-chain-sphchain}, Corollary \ref{kor-pokrivanje},
Proposition \ref{prop-proper}, Lemma \ref{l-zeta}, Proposition
\ref{pr-S-J}(1) and the fact that  $\fmesh$ is a computable
function. The fact that $\Omega $ is c.e.\ and the fact that for
each $k\in \mathbb{N}$ there exists $l\in \mathbb{N}$ such that
$(k,l)\in \Omega $ imply that there exists a computable function
$g:\mathbb{N}\rightarrow \mathbb{N}$ such that $(k,g(k))\in \Omega
$ for each $k\in \mathbb{N}$.

Let $k\in \mathbb{N}$. By Proposition \ref{sfera-2} for each $x\in
\bigcup (\partial \mathcal{H}_{g(k)})$ there exists $y\in S$ such
that $d(x,y)<3\cdot 2^{-k}$. Now Lemma \ref{l-zeta} and Lemma
\ref{aproks} imply that $S$ is computable. \qed


\begin{thm}\label{thm-2} Let $(X,d,\alpha )$ be a computable metric space which is locally
computable. Let $E$ be an $n-$cell in $(X,d)$ and suppose $E$ and
the boundary sphere of $E$ are co-c.e.\ in $(X,d,\alpha )$. Then
$E$ is computable.
\end{thm}
\proof We proceed in a similar way as in the proof of Theorem
\ref{thm-1}. First, we may assume that $(X,d,\alpha )$ has compact
closed balls and the effective covering property. Let $f:I^{n}
\rightarrow E$ be a homeomorphism. Let $S=f(\partial I^{n} )$.
Choose sets $W_{i}^{\rho }$, $1\leq i\leq n$, $\rho \in \{0,1\}$,
so that each of these sets is a finite union of rational balls and
so that the closures $\overline{W^{0}_{i}}$ and $\overline{W_{i}
^{1}}$ are disjoint and $f(A_{i} ^{\rho })\subseteq W_{i} ^{\rho
}$ for all $i\in \{1,\dots ,n\}$ and $\rho \in \{0,1\}$. Let
$k_{0} \in \mathbb{N}$ be such that $2\cdot 2^{-k_{0} }
<d(W^{0}_{i},W_{i} ^{1})$
 for each $i\in
\{1,\dots ,n\}$ (such $k_{0} $ certainly exists since
$\overline{W^{0}_{i}}$ and $\overline{W_{i} ^{1}}$ are compact and
disjoint for each $i\in \{1,\dots ,n\}$).

Using Proposition \ref{kugla-1}, Lemma \ref{lema-nova} and
Proposition \ref{fdiam} we conclude that for each $k\in
\mathbb{N}$ there exists $l\in \mathbb{N}$ with the following
properties:
\begin{equation}\label{2tm1}
\widehat{\mathcal{H}}_{l}\mbox{ is an
}n\mbox{-chain},~\mathcal{H}_{l}\mbox{ covers }E,~\partial
\mathcal{H}_{l}\mbox{ covers }S,
\end{equation}
\begin{equation}\label{2tm3}
\fmesh(l)<2^{-(k+k_{0} )},~\mathcal{H}_{l}\mbox{ is }2^{-(k+k_{0}
)}-\mbox{ proper}
\end{equation}
and
\begin{equation}\label{2tm5}
\bigcup \left(\partial _{i} ^{\rho
}\widehat{\mathcal{H}}_{l}\right)\subseteq W_{i} ^{\rho }
\end{equation}
for all $i\in \{1,\dots ,n\}$ and $\rho \in \{0,1\}$.

As in the proof of Theorem \ref{thm-1} we conclude that there
exists a computable function $g:\mathbb{N}\rightarrow \mathbb{N}$
such that (\ref{2tm1}), (\ref{2tm3}) and (\ref{2tm5}) hold for
each $k\in \mathbb{N}$ and $l=g(k)$. Let $k\in \mathbb{N}$. By
Proposition \ref{kugla-2} for each $x\in \bigcup
\mathcal{H}_{g(k)}$ there exists $y\in E$ such that $d(x,y)<7\cdot
2^{-k}$ and therefore $E$ is computable. \qed

Let us mention that a computable $n$-cell  need not be computably
homeomorphic to the unit ball in $\mathbb{R}^{n} $. It has been
shown in \cite{mi:mib} that there exists a computable arc $E$ in
$\mathbb{R}^{2}$ with computable endpoints, but such that there
exists no homomorphism $[0,1]\rightarrow E$ which is a computable
function. Similarly, a computable $(n-1)$-sphere need not be
computably homeomorphic to the unit sphere in $\mathbb{R}^{n} $
(\cite{mi:mib}).

\section{Conclusion}

In this paper we have seen that topology plays an important role
regarding the computability of co-c.e.\ sets in computable metric
space. We have seen that the topological types of an arbitrary
dimensional sphere and an arbitrary dimensional cell behave well
from this viewpoint not just in Euclidean space but in any
computable metric space which is locally computable, in particular
in any computable metric space which has the effective covering
property and which is locally compact. Such a computable metric
space is for example the Hilbert cube $I^{\infty}$, equipped with
a natural computability structure (see e.g.\ \cite{zi:mib}).

It should be mentioned that co-c.e.\ spheres, as well as co-c.e.\
cells with co-c.e.\ boundary spheres, need not be computable in a
computable metric space which is not locally computable. Moreover,
by \cite{zi2:mib}, there are examples of computable metric spaces
$X$ and $Y$ such that $X$ has the effective covering property and
$Y$ is compact, but such that both $X$ and $Y$ have noncomputable
co-c.e.\ topological circles and a noncomputable co-c.e.\ arcs
with computable endpoints.



\section*{Acknowledgements}
The author would like to thank the anonymous referees for their
careful work and many useful suggestions and corrections.
Furthermore, the author is grateful to Professor Sibe
Marde\v{s}i\'{c} and Professor Ivan Ivan\v{s}i\'{c} for their
helpful comments.

\vspace{-20 pt}

\end{document}